\documentclass[12pt]{iopart}

\usepackage{iopams}
\usepackage{amssymb}
\usepackage{setstack}
\usepackage{graphicx}
\usepackage{dcolumn}
\usepackage{bm}
\usepackage{dsfont}
\usepackage{amssymb}

\newcommand{\atanh  }{{\rm{atanh}}}

\newcommand{\bh     }{\mbox{\boldmath$h$}}

\newcommand{\bzero  }{\mbox{\boldmath$0$}}
\newcommand{\bx     }{\mbox{\boldmath$x$}}

\usepackage{color}

\begin{document}
\date{\today}
\title{Gallager error correcting codes for binary asymmetric channels}
\author{I. Neri, N. S. Skantzos and D. Boll\'e}
\address{Instituut voor Theoretische Fysica, Katholieke Universiteit Leuven, Celestijnenlaan 200D, B-3001 Leuven, Belgium}
\ead{izaak.neri@fys.kuleuven.be, nikos@itf.fys.kuleuven.ac.be, desire.bolle@fys.kuleuven.be}
 \begin{abstract}
We derive critical noise levels for Gallager codes on asymmetric
channels as a function of the input bias and the temperature.  Using a statistical mechanics approach we study the
space of codewords and the entropy in the various decoding
regimes.   We further discuss the relation of the 
convergence of the messsage passing algorithm with the endogeny property and complexity, characterizing solutions of
recursive equations of distributions for cavity fields.
 \end{abstract}

\section{Introduction}

Error-correcting codes play a central role in modern communication.
These are used to communicate reliably in noisy media such as
satellite and mobile communication. Currently, there is a wide range
of error-correcting schemes, ranging from the classic Reed-Solomon
codes \cite{reed-solomon}, used today in mass storage media, to the
more recent turbo codes \cite{turbo} and low-density parity-check
(LDPC) codes \cite{gallager1,gallager2} that have both shown near
optimal performance.

Error-correcting codes exploit the idea of introducing redundancy
into the message. The extra `redundant' bits are constructed in a
way known to both sender and receiver by correlating the message
bits. If the channel noise is not too high the receiver can
successfully decode and retrieve the exact original message. To
minimize the transmission costs the amount of redundancy must be as
small as possible. This inevitably makes the code more prone to
errors; any good error correcting scheme must minimize both the
required redundancy and the error probability. It is not at all
a priory clear what are the limits in this trade-off. It was
in 1948 that Claude Shannon blazed the trail \cite{shannon1948} and
proved that good error-correcting codes  with arbitrarily small
error probabilities do exist as long as the amount of redundancy is not smaller than a certain level, the so-called channel
capacity.

However, Shannon's groundbreaking proof was not suggestive as to how
to construct practical useful codes that reach the Shannon limit. In 1962 Gallager proposed the family of the
so-called low-density parity-check codes \cite{gallager1,gallager2}.
Although conceptually simple, this coding scheme was largely
forgotten due to the computational limitations of the time.
Currently, however, with the advent of the computer era, they are
recognized as one the best schemes available.  The LDPC are easy to construct, have a low complexity and perform near the Shannon limit \cite{Rich2003}.

Gallager codes were re-discovered by MacKay and Neal \cite{MN} at
around the same time when Sourlas \cite{sourlas} showed that
statistical physics can be used to estimate the performance of error
correcting codes.  These two events brought in a surge of activity as
well as an influx of research ideas from physics to information
theory and vice versa. In particular, from a physics viewpoint,
error-correcting codes have been so far studied quite extensively.
For example, low-density parity-check ones on  binary symmetric
channels \cite{Vic1999, montanari,franz,kabashima2,murayama}, on real-valued
channels \cite{skantzos1,tanaka}, on irregular graphs \cite{vicente}
while more recently the error exponent was calculated in
\cite{skantzos2,rivoire}. Turbo codes have been studied in
\cite{montanari1,montanari2}. For a more complete review on the
subject see \cite{saad_review, Mont2006}. Clearly, the bibliography of
statistical physics of codes is largely biased towards the
low-density parity-check ones: this is because the recently developed finite-connectivity techniques offer an ideal
toolbox for the theoretical study of this field.  Currently, the
more recent developments in the physics of finitely connected
systems allow one to extend previous results with algorithms that perform better
\cite{Migl2005,hatchett,wemmenhove}.  Altough these algorithms reach the computational limits of today, maybe one day they will also become useful.

Qualitatively speaking the emerging picture for Gallager codes is
that for sufficiently small noise levels, decoding with a message-passing algorithm with a linear computational complexity in the block size is possible and
the error-free state is the only stable state. For higher noise
levels, one finds a transition to a regime where suboptimal states
are created (marking the so-called spinodal or dynamical transition)
and where the message passing algorithms fail to find the
most probable solution. For higher noise levels, a second transition
occurs (thermodynamic transition) where the error-free solution
ceases to be dominant. This marks the upper theoretical bound for
error-free communication.  This means that block-wise maximum likelihood decoding, which is shown to be NP-complete \cite{Berle1978}, fails.

In this paper we study Gallager codes on the family of binary
asymmetric channels (BAC). The two extreme cases of this family include
the binary symmetric channel (which has been the key actor in nearly
all previous research) and the (fully asymmetric) Z-channel. The
latter is used in communications through optical fibers. Within
replica symmetry we calculate the location of the static and dynamic
transitions. We also present phase diagrams describing
where the frozen phase and clustered phases appear. As a reference point to
test our theory we have used known results from information theory
\cite{Wang2005} and shown that it reproduces them with very good
agreement.

Our paper is organized as follows: In the following section we
provide the model definitions for Gallager codes and the
decoding process. In section \ref{sec:statmech} we set up the
decoding problem in statistical mechanical terms. In section
\ref{sec:rcm} we derive the thermodynamic quantities for a simple
limiting case (dense codes) while in section \ref{sec:f} we compute
the free energy of the binary asymmetric channel. In section
\ref{sec:bp} we discuss the failure of belief propagation (BP) while in
section \ref{sec:1rsb} we present results from a one-step replica
symmetry-breaking scheme. We end this paper with a discussion in
section~\ref{sec:disc}.

\section{Model definitions}
\label{sec:defs}
\subsection{Gallager codes}
The aim is to send a message reliably through a noisy medium.  Hereby four processes are of importance: the generation of the message by the source, the encoding process, the noise and the decoding process.

We consider a source which
produces messages $\bsigma^0\in\{-1,1\}^N$ with probability
\begin{equation}
P_{\rm in}(\bsigma^0) = \prod_{i=1}^N P_{\rm in}(\sigma_i^0;b) =
\prod_{i=1}^N b\delta_{\sigma_i^0,1}+(1-b)\delta_{\sigma^0_i,-1} \:,
\end{equation}
with $b\in[0,1]$ the bias of the input signal.

\begin{figure}
\begin{center}
\includegraphics[width=0.3 \textwidth]{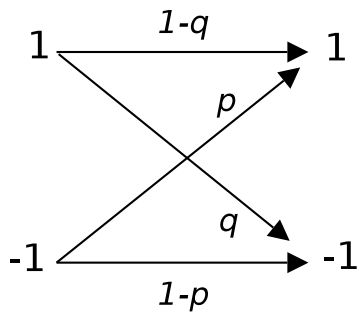}
\caption{A graphical representation of the binary asymmetric
channel: noise corrupts the different bits with a different
probability.}\label{fig:channel}
\end{center}
\end{figure}

The message is sent from one point to another through a noisy
channel. To communicate the message in an error-free way redundant
bits are added to the message before it is sent through the
channel (encoding process).  The encoding process
is defined by the map $\mathcal{G}:\{-1,1\}^N\rightarrow
\{-1,1\}^M: \bsigma^{0} \rightarrow \bsigma$, with $N<M$.   The
elements of the image $\mathcal{C}$ of $\mathcal{G}$ are called the
codewords. Shannon, in his original paper \cite{shannon1948}, showed
that for a family of codes, having a completely random set of
codewords, it is possible for $N\rightarrow \infty$ to decode
errorlessly with probability one as long as the code rate $R =
\frac{N}{M}h(b)$, with $h(b)=-b\log_2b-(1-b)\log_2(1-b)$ the binary
entropy, is smaller or equal to the maximal admissible amount of
information we can send through the channel.  This is given by the
so-called channel capacity $\mathtt{C}$ (see
\ref{app:capBAC} for a computation of the channel capacity for the
BAC). However, as Shannon's random encoding turns out to be
inefficient for practical error correction a new encoding/decoding
strategy was sought. Gallager, among others, proposed a scheme for
introducing more structure in the set of codewords
\cite{gallager1,gallager2}. In particular, he suggested the linear
space of codewords:
\begin{eqnarray}
 \mathcal{C} &=& \Big\{\bsigma \in \{-1,1\}^M |\mathbb{H}\ast \bsigma = \mathbf{1}\Big\}\:,
 \label{eq:codeword_set}
\end{eqnarray}
with
\begin{eqnarray}
 \left(\mathbb{H} \ast  \bsigma\right)_i \equiv \prod^M_{j=1} \sigma^{\mathbb{H}_{ij}}_j \:,&\quad \forall i=1,  \cdots, M-N\:.
\end{eqnarray}
$\mathbb{H} = \Big[ \mathbb{C}_1 | \mathbb{C}_2 \Big]$ is the
parity check matrix which is a sparse $(M-N)\times M$ matrix
with elements $\mathbb{H}_{ij}\in\{0,1\}$. The symbol
$\Big[\mathbb{C}_1 | \mathbb{C}_2 \Big]$ denotes concatenation of
two matrices. The matrix $\mathbb{C}_1$ is of dimension $(M-N)\times
N$ and $\mathbb{C}_2$ is an invertible matrix of dimension
$(M-N)\times (M-N)$. The elements of $\mathbb{C}_1$ and
$\mathbb{C}_2$ lie in $\left\{0,1\right\}$.  In regular
Gallager codes the parity check matrix is constructed such that
there are $K$ non-zero elements per row and $C$ non-zero elements
per column. In irregular codes the number of ones per
row and per columns are drawn from a distribution. Counting the
number of ones in this matrix provides for regular codes the
relation $R=[1-{C}/{K}]h(b)$ which expresses the code rate
in terms of the code parameters. The $M-N$ equations implied by
(\ref{eq:codeword_set}) are the parity-check equations.  Using Gaussian elimination one can bring $\mathbb{H}$ to a
systematic form described by
$\mathbb{A}=\left[\mathbb{P}\,\Big|\,\mathds{1}_{M-N}\right]$ with
$\mathbb{P}=\mathbb{C}_2^{-1}\mathbb{C}_1 $ such that
$\mathbb{H}=\mathbb{C}_2\mathbb{A}$ and where $\mathds{1}_\ell$ is
the $\ell\times \ell$ identity matrix. The matrices $\mathbb{H}$ and
$\mathbb{A}$ span the same space and are thus equivalent parity
check matrices. We can now define the generator matrix
$\mathbb{G} = \left[\frac{\mathds{1}_N }{\mathbb{P}}\right]$ such
that due to the mod-2 arithmetic one obtains
$\mathbb{H}\mathbb{G}=[\mathbb{C}_1|\mathbb{C}_2]\
[\mathds{1}_N |
\mathbb{C}_2^{-1}\mathbb{C}_1]^{\dagger}=\mathbb{C}_1+\mathbb{C}_1= \bzero$.
With these definitions encoding is realized through $\bsigma =
\mathbb{G} \ast \bsigma^0$. This implies that $\sigma_i=\sigma^0_i$
for $i=1,\ldots,N$.

Channel noise can be seen as a bit-flipping operation. The effect of noise can be presented as a transformation $\bsigma\to
\brho = (\nu^0_1\sigma_1,\ldots,\nu^0_M\sigma_M)$, where
$\bnu^0\in\{-1,1\}^M$ represents the channel true noise
vector.  The channel can be represented by the probability $P_{\rm chan}(\bnu^0|\bsigma)$ of a true noise
vector given the message.  For the BAC, $P_{\rm chan}(\bnu^0|\bsigma)$ equals (figure \ref{fig:channel}),
\begin{eqnarray}
P_{\rm chan}(\bnu^0|\bsigma) &=& \prod^M_{i=1}P_{\rm chan}(\nu_i^0|\sigma_i) \:,
\label{eq:posterior}
\end{eqnarray}
with
\begin{eqnarray}
 \fl
 P_{\rm chan}(\nu^0|\sigma) = (1-p)\delta_{\sigma,-1}\delta_{\nu^{0}, 1} + p \delta_{\sigma,-1}\delta_{\nu^{0}, -1} + q \delta_{\sigma,1}\delta_{\nu^{0},-1} +
 (1-q)\delta_{\sigma,1}\delta_{\nu^{0},1}\:.
\end{eqnarray}
The parameters $p,q\in[0,1]$ give the bit-flip probabilities of the
channel.  In (\ref{eq:posterior}) we assumed that the channel is
memoryless. For convenience we define the variable $\kappa =
p/q\in[0,1]$ such that the binary symmetric channel corresponds to
$\kappa = 1$ while for $\kappa = 0$ one obtains the fully asymmetric
Z-channel, which is of interest for optical communication (with the
two states representing the presence or absence of light in the
channel).

The receiver at the other end of the
channel uses a prescribed set of operations to extract the original message from the
received word (decoding). 
After obtaining the bit stream $\brho$ the receiver is required to
solve, using the aforementioned properties of the generator matrix,
the equations $\mathbb{H}\ast\brho=\mathbb{H}\ast\bnu$.  Among the solutions of these equations an
estimate $\hat{\bnu}$ for the true noise $\bnu^0$ is obtained. Once this is
found an estimate for the original message $\hat{\bsigma}^0$
immediately follows.
The estimates of the single bits are obtained by calculating the single bit marginals
\begin{eqnarray}
 P_i &=& \sum_{\bnu\setminus \nu_i}P_{\rm dec}(\bnu|\brho, \mathbb{H}) = \sum_{\bnu\setminus \nu_i,\: \bnu \in \mathcal{C}}P_{\rm dec}(\bnu|\brho)\:. \label{eq:marginals}
\end{eqnarray}
The notation $\bnu\setminus \nu_i$ denotes the set of components of
$\bnu$ excluding the $i$-th. The choice of $P_{\rm dec}(\bnu|\brho)$
determines the decoding process while $\bnu$ are the variables
of this decoding process.  At first sight it seems impossible to
calculate these quantities as we need $2^{M-1}$ operations. However,
LDPC codes owe their success in the existence of a
belief-propagation algorithm  \cite{pearl} (see also
\cite{Vic1999,yedidia}), whose computational complexity scales
linearly in the system size $M$, able to calculate the above
marginals. This is achieved by interpreting the $M-N$ parity check
equations of (\ref{eq:codeword_set}) as a bipartite graph (the
so-called Tanner graph) in which $M$ variable nodes,
associated to each $\nu_i$, are connected to $M-N$ check
nodes associated to each of the constraints of
(\ref{eq:codeword_set}).

The performance of the code can be determined through a loss
function \cite{Iba1999}. If we take as loss function $L(\bnu,
\bnu^0) = -\sum^M_{i=1}\nu_i\nu^0_i$, which is the overlap between
the true noise vector and the variables of the decoding
process, the optimal estimator can be shown to be given by
$\hat{\nu}_i={\rm sign}\left(\sum_{\nu_i}P_i\nu_i\right)\equiv
\rm sign\langle\nu_i\rangle$ \cite{Nish2001}. Thus we measure
the performance through the order parameter $\rho$ defined as
\begin{eqnarray}
 \rho &\equiv& \frac{1}{M}\sum^M_{i=1}\overline{ \rm sign\langle \nu_i\nu^0_i\rangle} \:. \label{eq:order}
\end{eqnarray}
The brackets denote the average over (\ref{eq:marginals}) and the
bar denotes the average over $\brho$ and $\mathbb{H}$.

\subsection{Decoding processes}
Without loss of generality we can represent the conditional
probability $P_{\rm dec}(\bnu|\brho)$ through
\begin{eqnarray}
 P_{\rm dec}(\bnu|\brho) &=&\mathcal{N}(\brho) \prod^M_{i=1} \exp\left(\nu_i\beta_1H_1\right) \delta_{\rho_i, 1} +
 \exp\left(\nu_i\beta_{-1} H_{-1}\right)\delta_{\rho_i, -1} \:,\label{eq:decode}
\end{eqnarray}
distinguishing between different states for each received bit. The subindex corresponds with the value of the received bit $\rho_i$.  The
normalization constant $\mathcal{N}(\brho)$ is independent of the
decoding variables and will be left out.  This will be important for
the calculation of the entropy. 
 The parameters $ H_1$ and $H_{-1}$,
also called the Nishimori parameters, determine the decoding scheme
in the case of symbol-wise maximum a-posteriori probability (symbol-wise MAP). In symbol-wise MAP we want to choose the probability distribution $P_{\rm dec}(\bnu|\brho)$ such that $\rho$ is maximal. Following \cite{Iba1999} we can find these
parameters by identifying (\ref{eq:decode}) with the true posterior
probability distribution $P_{\rm post}(\bnu|\brho)$ determined by the characteristics of the
source and the channel noise.  Using Bayes' rule we obtain
\begin{eqnarray}
 P_{\rm post}(\nu_i|\rho_i)
&=& \frac{\left(\sum_{\sigma_i}P(\rho_i|\sigma_i, \nu_i)P(\sigma_i|\nu_i)\right)P(\nu_i)}{P(\rho_i)}\:,
\end{eqnarray}
and
\begin{eqnarray}
 P(\sigma_i|\nu_i) &=& \frac{P_{\rm chan}(\nu_i|\sigma_i)P_{\rm prior}(\sigma_i)}{P(\nu_i)}\:.
\end{eqnarray}
One can easily write down the probabilities $P_{\rm chan}(\nu_i|\sigma_i)$ and
$P(\rho_i|\sigma_i,\nu_i)$ from the channel description of figure
\ref{fig:channel}. For instance
$P(\rho_i|\nu_i,\sigma_i)=\sum_{\sigma,\rho=\pm1}\delta_{\sigma_i,\sigma}\delta_{\rho_i,\rho}\delta_{\nu_i,\sigma\rho}$.
 For the a priory probability of codewords we have
\begin{eqnarray}
  P_{\rm prior}(\bsigma) &=& \frac{\delta\left(\mathbb{H}\ast \bsigma = \mathds{1}\right)\prod^{N}_{i=1}P_{\rm in}(\sigma_i)}{\sum_{\bsigma}\delta\left(\mathbb{H}\ast \bsigma = \mathds{1}\right)\prod^{N}_{i=1}P_{\rm in}(\sigma_i)}\:, \label{eq:PriorSigma}
\end{eqnarray}
since the first $N$ bits of the codeword are copies of the
original message and all codewords must satisfy
(\ref{eq:codeword_set}). From $P_{\rm post}(\bnu|\brho)= P_{\rm
dec}(\bnu|\brho)$ we then find that the Nishimori parameters become
\begin{eqnarray}
\beta_1 = 1\:, &\quad & H_1(b) =  \frac{1}{2}\log\frac{(1-q)b}{p(1-b)}\:,\\
\beta_{-1} = 1\:, & \quad & H_{-1}(b) = \frac{1}{2}\log\frac{(1-p)(1-b)}{qb} \:.
\end{eqnarray}
In general we will consider decoding processes where $\beta_1 = \beta_{-1} = \beta$.  When $\beta\rightarrow \infty$ we get block-wise MAP decoding.  We remark that for unbiased channels symbol-wise and block-wise MAP decoders perform the same as symbol-wise and block-wise maximum likelihood decoders.

\section{Statistical mechanics for Gallager codes}
\label{sec:statmech}

\subsection{The partition function}

To begin the statistical mechanical analysis of the decoding
process we define the equilibrium Boltzmann measure of candidate
noise vectors given the parity check matrix, the true noise vector
and the received bit stream:
\begin{equation}
p_{\rm equil}(\bnu|\mathbb{H},\bnu^0,\brho)
=\frac{1}{\mathcal{Z}(\mathbb{H},\bnu^0,\brho)}
\frac{1}{\mathcal{N}\left(\brho\right)}\delta\left[\mathbb{H} \ast
\bnu = \mathbb{H} \ast \bnu^0 \right]\,P_{\rm dec}(\bnu|\brho)\:,
\end{equation}
where $\mathcal{Z}(\mathbb{H},\bnu^0,\brho)$ represents the partition function of our system:
\begin{eqnarray}
\mathcal{Z}(\mathbb{H},\bnu^0,\brho) =\frac{1}{\mathcal{N}\left(\brho\right)}\sum_{\bnu} \delta\left[\mathbb{H} \ast \bnu = \mathbb{H} \ast \bnu^0 \right]\,P_{\rm dec}(\bnu|\brho)\:.
\end{eqnarray}
To find the behavior of $\rho$ in (\ref{eq:order}),
we calculate the typical value $f_t$ of the free energy $f = -\frac{1}{\beta M}\log \mathcal{Z}$ for $M\rightarrow \infty$.  Assuming self-averaging we can find $f_t$ by calculating the code- and noise-averaged free energy $\overline{f}$, given by
\begin{eqnarray}
 \overline{f} &=& -\lim_{M\to\infty}\frac{1}{\beta M}\sum_{\mathds{H}}P(\mathds{H})\sum_{\brho\bnu^0}p_{\rm post}(\bnu^0,\brho| \mathbb{H})\log \mathcal{Z}(\mathds{H}, \bnu^0,\brho) \:. \label{eq:freeQ}
\end{eqnarray}
The hardcore restriction, which imposes that candidate noise vectors
must satisfy the parity checks, can be written as
\begin{eqnarray}
 \fl \delta\left[\mathbb{H}\ast\bnu = \mathbb{H} \ast \bnu^0 \right] = \lim_{\gamma\rightarrow \infty}\exp\left[\gamma \sum_{\langle j_1, j_2, \cdots, j_K\rangle}\mathcal{T}_{\langle j_1, j_2, \cdots, j_K\rangle }\left(J_{j_1j_2\cdots j_K}\nu_{j_1}\nu_{j_2}\cdots \nu_{j_K}-1\right)\right]\:,
\end{eqnarray}
with $J_{j_1j_2\cdots j_K} = \nu^0_{j_1}\nu^0_{j_2}\cdots
\nu^0_{j_K}$ and
\begin{eqnarray}
  \mathcal{T}_{\langle j_1, j_2, \cdots, j_K\rangle }  &=&
  \left\{\begin{array}{ccl}1 &{\rm if} & \prod^K_{l=1}\mathbb{H}_{ij_l} = 1  \ {\rm for \ some} \ \ i \in
\{1,\ldots,M-N\} \\ 0 &{\rm if} & {\rm otherwise} \end{array}\right. \:.
\end{eqnarray}
The probability distribution of the tensor $\mathcal{T}$ follows from the
statistics of $\mathbb{H}$, namely
\begin{eqnarray}
 \fl P(\mathcal{T}) = \frac{1}{\mathcal{M}}\prod_{\langle j_1,j_2, \cdots, j_K \rangle}\left[C\frac{(K-1)!}{M^{K-1}}\delta\left[ \mathcal{T}_{\langle j_1,j_2, \cdots, j_K \rangle}-1\right] + \left[1-C\frac{(K-1)!}{M^{K-1}}\right]\delta( \mathcal{T}_{\langle j_1,j_2, \cdots, j_K \rangle})\right]
\nonumber \\
\times  \prod^{M}_{l=1}\delta\left(\sum_{\langle j_2,\cdots, j_K \rangle; j_1 =
l} \mathcal{T}_{\langle j_1,j_2, \cdots, j_K \rangle} - C\right)\:.
\label{eq:ProbA}
\end{eqnarray}
$\mathcal{M}$ is the normalization constant, i.e.\@ $ \mathcal{M} =
e^{-MC}(\frac{C^C}{C!})^M$. We have used the notation $\langle j_1,
j_2, \cdots, j_K\rangle$  to denote the ordered set
$j_1<j_2<\cdots<j_K $. The joint probability $ p_{\rm post}(\bnu^0,\brho| \mathbb{H})$ in
(\ref{eq:freeQ}) does not factorize due to the asymmetry of the
channel. It can be evaluated through
\begin{eqnarray}
 p_{\rm post}(\bnu^0,\brho|\mathbb{H})&=&\sum_{\bsigma}P_{\rm prior}(\bsigma)P_{\rm chan}(\bnu^0|\bsigma)\delta[\brho\bnu^0,\bsigma] \:.
\end{eqnarray}
After making the gauge transformation $\nu_i\rightarrow
\nu_i\nu^0_i$, we have the following partition function
\begin{eqnarray}
\fl \mathcal{Z}(\left\{h_i\right\},\mathbb{H}) &=& \sum_{\bnu}\exp\left[
\gamma\sum_{\langle j_1,j_2, \cdots, j_K \rangle}
\mathcal{T}_{\langle j_1,j_2, \cdots, j_K
\rangle}\left(\nu_{j_1}\nu_{j_2}\cdots \nu_{j_K}-1\right) + \beta \sum^M_{i=1}h_i\nu_i
\right] \:, \label{eq:PartSum}
\end{eqnarray}
modulo irrelevant multiplicative constants. The quenched fields
$h_i$  are drawn from the distribution
 \begin{eqnarray}
 \mathcal{P}\left( \left\{h_i\right\}\right) &=& \frac{\sum_{\bsigma}\delta\left(\mathbb{H}\ast \bsigma = \mathbf{1}\right)\prod^N_{i=1}p_{b}(h_i, \sigma_i) \prod^M_{i=N+1}p_{\frac{1}{2}}(h_i, \sigma_i)}{\sum_{\bsigma}\delta\left(\mathbb{H} \ast \bsigma = \mathbf{1}\right)\prod^N_{i=1}p_{b}(\sigma_i)\prod^M_{i=N+1}p_{\frac{1}{2}}(\sigma_i)}\:, \label{eq:PFields}
\end{eqnarray}
with
\begin{eqnarray}
\fl p_b(h, \sigma) =(1-q)b\delta\left(\sigma, 1\right)\delta\Big(h-H_1(b)\Big) + p(1-b)\delta\left(\sigma, -1\right)\delta\Big(h+H_1(b) \Big)
\nonumber \\
  + (1-p)(1-b)\delta\left(\sigma, -1\right)\delta\Big(h-H_{-1}(b)  \Big) + qb\delta\left(\sigma, 1\right)\delta\Big(h+H_{-1}(b) \Big)\:, \label{eq:BAChdistri2}
\end{eqnarray}
and $p_b(\sigma)= \int dh p_b(h, \sigma)$.

\subsection{Gauge transformation}

The gauge theory of disordered systems, pioneered by Nishimori
\cite{Nish1981}, uses symmetry relations to derive a number of exact
results. Of particular interest is the Nishimori line on which one
can compute exactly the internal energy and one can show that there
are no replica symmetry breaking effects. For error correcting
codes, using symbol-wise MAP decoding with $\beta=1$ turns out to be equivalent
to computing decoding observables on the Nishimori line.

For an unbiased BSC we have $p_{\frac{1}{2}}(h, 1) =
p_{\frac12}(h, -1)$. This model falls then in the category of
channels characterized in \cite{montanari}. Since the above
distribution (\ref{eq:BAChdistri2}) fullfils the conditions
$p_b(-h,-\sigma)=e^{-2h}p_b(h,\sigma)$ we can write the free energy
in a more symmetric form as in \cite{montanari}. For any observable
$\mathcal{O}(h)$ we can write
\begin{eqnarray}
\int^{+\infty}_{-\infty}p(h_i,\sigma_i)\mathcal{O}\left(h_i\right) = \int^{+\infty}_0 dh_i \sum_{\tau_i}\rho\left(h_i,\sigma_i\tau_i\right)e^{h_i\tau_i}\mathcal{O}\left(h_i\tau_i\right)\:,
\end{eqnarray}
with
\begin{eqnarray}
\rho\left(h_i,\sigma_i\right) = \frac{p_b(h_i,\sigma_i) + p_b(-h_i,-\sigma_i)}{2\cosh\left(h_i\right)}\:.
\end{eqnarray}
Making the transformations $\sigma_i\rightarrow \sigma_i\tau_i$
and also $(\tau_i,\nu_i)\rightarrow (\tau_i\mu_i, \nu_i\mu_i)$,
with $\delta(\mathbb{H}\ast \bmu) = 1$, we arrive at the more
symmetric form
\begin{eqnarray}
\fl -\beta M\overline{f}(\mathbb{H}) \sim \sum_{\bsigma}\int^{\infty}_0\prod^M_{i=1}dh_i\rho(h_i,\sigma_i)
\sum_{ \btau}\delta\left(\mathbb{H}\ast \bkappa\right)
\nonumber \\
\times
\sum_{\bmu}\delta\left(\mathbb{H}\ast \bmu\right) \exp\left[\sum^M_{i=1}h_i\tau_i\mu_i\right]\log\left(\sum_{\bnu}\delta\left(\mathbb{H}\ast \bnu\right)\exp\left[\beta \sum^M_{i=1}h_i\nu_i\tau_i\right]\right) \:, \nonumber \\ 
\end{eqnarray}
with $\bkappa=(\sigma_1\tau_1,\ldots,\sigma_M\tau_M)$. At $\beta = 1$ we can, using the
techniques in \cite{Nish2001}, exploit this symmetry to prove that
the thermodynamic state is replica symmetric. The energy $\epsilon_{\beta} = \partial_{\beta}\beta f$ at $\beta=1$ equals
\begin{eqnarray}
\fl\epsilon_{\beta=1} &= -\int \prod^M_{i=1}dh_i\mathcal{P}\left(\left\{h_i\right\} \right)\left(\frac{\sum^M_{i=1}h_i}{M}\right)
\nonumber \\
&= - \frac{\int da\ p(a;b)\
\sum_{\bsigma}\delta\left(\mathbb{H}\ast \bsigma\right)
\prod^N_{i=1}p_a(\sigma_i)\langle h
\rangle_{h|\sigma_1,a}}{\sum_{\bsigma}\delta\left(\mathbb{H}\ast
\bsigma\right)
\prod^N_{i=1}p_b(\sigma_i)\prod^M_{i=N+1}p_{\frac12}(\sigma_i)}
\label{eq:energyFerro}\:.
\end{eqnarray}
The distribution $p(a;b)$ is defined as
\begin{eqnarray}
 p(a;b) &=& \frac{N}{M}\delta(a-b) + \frac{M-N}{M}\delta(a-\frac12)\:.
\end{eqnarray}
The average $\langle \cdots\rangle_{h|\sigma,a}$ is over
$p_a(h|\sigma)$. One can also prove that $ \rho_{\beta=1}\geq
\rho_{\beta}$ which is equivalent to the statement that $\beta = 1$
corresponds with MPM decoding \cite{Nish1993}.

\section{A simple solvable detour: The random codeword model}
\label{sec:rcm}

\begin{figure}
\hfill
\begin{minipage}{.45\textwidth}
\begin{center}
\includegraphics[angle=-90, width = 1 \textwidth]{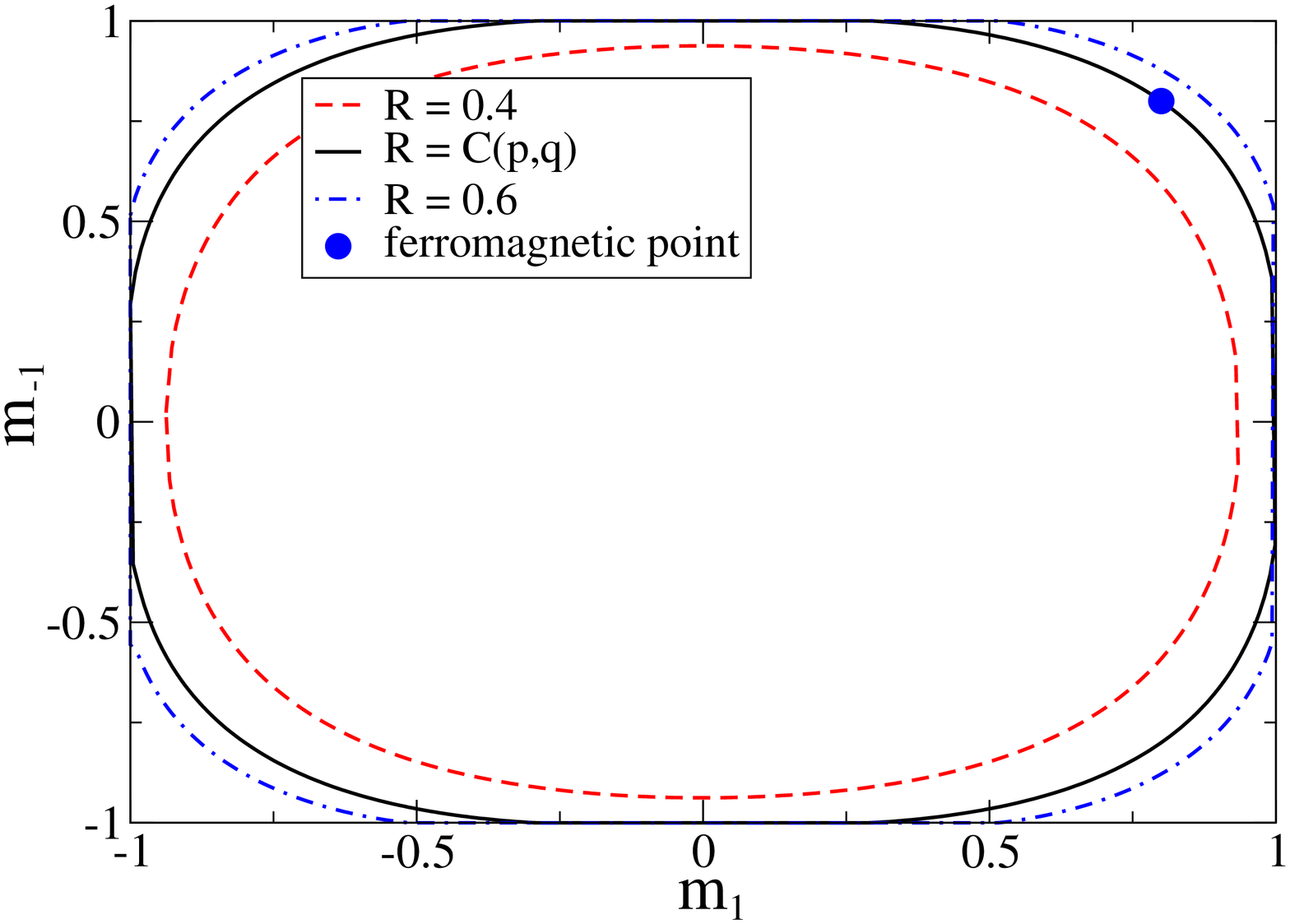}
\end{center}
\end{minipage}
\hfill
\begin{minipage}{.45\textwidth}
\begin{center}
\includegraphics[angle=-90,width= 1 \textwidth]{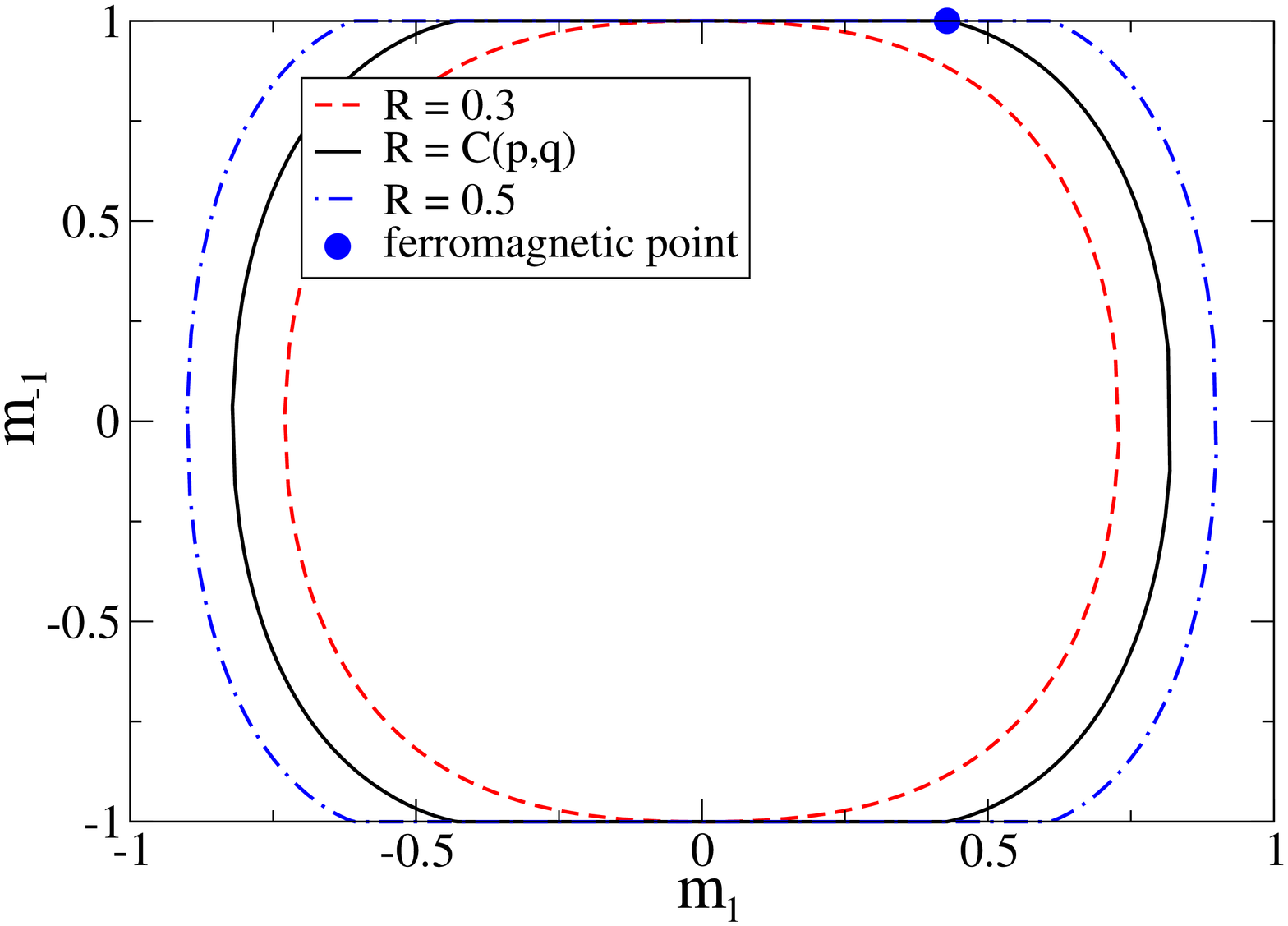}
\end{center}
\end{minipage}
\caption{The zero entropy lines, $s(m_1,m_{-1})=0$, for different
rates $R$ at $q=0$ and $p=0.4$ in the unbiased case of
$b=\frac{1}{2}$. For rates $R>$\texttt{C}$(p,q)$ the entropy is positive
and decoding is not possible. Left: BSC. Right: Z-channel}
\label{fig:shannon}
\end{figure}

\begin{figure}
\begin{center}
\includegraphics[angle=-90, width = 0.6 \textwidth]{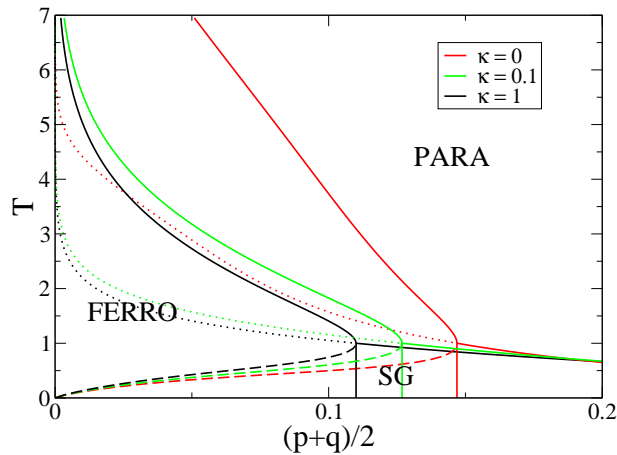}
\end{center}
\caption{The (T,(p+q)/2)-phase diagram for the random codeword
model with a rate $R=1/2$ for different degrees of the channel asymmetry
$\kappa=p/q$. Solid lines indicate the thermodynamic
transitions to the paramagnetic (PARA), spin glass (SG) or
ferromagnetic (FERRO) phases . The dotted line represents the
thermodynamic transition if freezing of the
paramagnetic solution is ignored.  The dashed line is the continuation of the
PARA-SG line.  } \label{fig:sRCMphase}
\end{figure}

Before we embark on the evaluation of the finitely connected case, we
consider the simple limiting case of
$K,C\to\infty$. This limit, implying an infinite number of parity
checks, is of course of small practical importance but nevertheless
very educational as it already contains a wealth of information
about the code's performance. It will also give us a first flavour
about the effects of asymmetry in Gallager codes. In this limit it
can be shown that the codewords $\bx\in\mathcal{C}$, for an unbiased source, are sampled
with a flat probability, thus this model is coined the `random
codeword model' (RCM).  These codewords determine the paramagnetic behavior of the system.  Besides these, the model also contains the ferromagnetic state $\bx^{(0)} = \bsigma$.   We choose $\bsigma= (1,1,\ldots,1)$.  We could say that this choice identifies $x_i = \nu_i\nu^0_i$, which corresponds with the analysis done before.
Below we follow the derivation as given in \cite{montanari}. The energies of the codewords, after the gauge transformation $x_i\rightarrow \rm sign(h_i)x_i$, are given by
\begin{eqnarray}
 \frac{E}{N} &=& \sum_{l=\pm1}\epsilon_l  = -\sum_{l=\pm1}|H_l|m_l \:,
\end{eqnarray}
with $N_l = \sum^M_{i=1}\delta(|h_i|,|H_l|)$ and $m_l = \frac{1}{N_l}\sum^M_{i=1}\delta(|h_i|,|H_l|)\sigma_i$.
The entropy of these states, for a given $m_1$ and $m_{-1}$, is equal to
\begin{eqnarray}
 \fl s\left(m_{1}, m_{-1}\right) = \left(R-1\right)\log 2 +  \left(\frac{1-q+p}{2}\right)Q(m_{1}) + \left(\frac{1+q-p}{2}\right)Q(m_{-1})\:,
\end{eqnarray}
with $Q(m)=-\sum_{\lambda=\pm 1}\frac12(1+\lambda m)\log[\frac12(1+\lambda m)]$.
The limit of maximum likelihood decoding is given by the noise levels $(p^*,q^*)$ where $s(m^F_1,m^F_{-1}) = 0$, with $(m^F_1,m^F_{-1})$ the magnetizations of the ferromagnetic state:
\begin{equation}
m^F_1={\rm sign}(H_1)\frac{1-q-p}{1-q+p}\:, \hspace{10mm}
m^F_{-1}={\rm sign}(H_{-1})\frac{1-q-p}{1+q-p}\:.
\end{equation}
  This zero entropy condition corresponds to $R=\mathcal{I}(p^*,q^*)$, with $\mathcal{I}$ the mutual information for an asymmetric channel, see equation (\ref{eq:mutualIas}).  We thus find the
Shannon limit back, see also figure \ref{fig:shannon}.  
In finite temperature decoding we restrict the energies $\epsilon_1$ and $\epsilon_{-1}$  by introducing the Lagrange parameters $\beta_1$ and $\beta_{-1}$.  The free energy $f(\beta_1,\beta_{-1})$ is defined through the Legendre transformation
\begin{eqnarray}
 f(\beta_1,\beta_{-1}) &=& s(\epsilon_1,\epsilon_{-1}) - \beta_1\epsilon_1 - \beta_{-1}\epsilon_{-1} \:.
\end{eqnarray}
We find that the entropy as a function of $\beta_1$ and $\beta_{-1}$, becomes zero when $(\beta_1,\beta_{-1}) = (\beta^f_1, \beta^f_{-1})$, with
\begin{eqnarray}
\fl
\lefteqn{\frac{1-q^*+p^*}{2}H\left[\frac{1-q^*-p^*}{1-q^*+p^*}\right]
+
\frac{1+q^*-p^*}{2}H\left[\frac{1-p^*-q^*}{1+p^*-q^*}\right]}
&&
\nonumber \\
&=&
\frac{1-q+p}{2}H\left[\frac{(1-q)^{\beta^f_1}-p^{\beta^f_1}}{(1-q)^{\beta^f_1}+p^{\beta^f_1}}\right]
+
\frac{1+q-p}{2}H\left[\frac{(1-p)^{\beta^f_{-1}}-q^{\beta^f_{-1}}}{(1-p)^{\beta^f_{-1}}+q^{\beta^f_{-1}}}\right]\:.
\end{eqnarray}
The entropy can become negative as a result of having a
partition sum dominated by atypical states. The number of these
states becomes zero when $M\rightarrow \infty$.  This corresponds
with an entropy crisis as found in the random energy model
\cite{Derrida}, \cite{Bouch1997}.  To avoid this we will introduce
the spin glass phase corresponding with the ground states of
the system.  The spin glass state has a free energy $f_{SG}$ given
by
\begin{eqnarray}
 f_{SG}(\beta_1,\beta_{-1}) &=& f_{P}(\beta^{f}_1,\beta^{f}_{-1}) \:,
\end{eqnarray}
with $s(\beta^{f}_1,\beta^{f}_{-1})=0$. The paramagnetic free energy $f_P$ is given by
\begin{eqnarray}
 \fl -f_{P}(\beta_1,\beta_{-1}) = R\log(2) + \left(\frac{1-q+p}{2}\right) \log\cosh \beta_1H_1 +  \left(\frac{1-p+q}{2}\right) \log\cosh \beta_{-1}H_{-1} \:. \nonumber \\
\end{eqnarray}
Comparing the free energies of the ferromagnetic, paramagnetic and spin glass state we find for $\beta_1=\beta_{-1}$ the phase diagram presented in figure \ref{fig:sRCMphase}.  We remark that increasing the degree of asymmetry in the
channel noise leads to a bigger ferromagnetic region.  The ferromagnetic-spin glass phase transition is given by $R=\mathcal{I}(p,q)$.  The triple point lies at $(\beta_1,\beta_{-1}) = (1,1)$.

\section{Free energy and saddle point equations}
\label{sec:f}

 In the more general case, the evaluation of the free energy
(\ref{eq:freeQ}) and of the various thermodynamic properties can be
done either with the replica
\cite{viana1985,kanter1986,WongSherr1987} or the cavity method
\cite{par1987,Mez2001}, both of which have been shown to lead to
identical results. Although the two methods differ in their
philosophy, they can be seen as two complementary sides of the same
coin, and together they can offer a more complete understanding of
the physics of the system under study. We follow here the replica
methodology and postpone our discussion on the cavity method for \ref{sec:cav}. The free energy is of the form
\begin{eqnarray}
\fl
 \overline{f} &=&  \lim_{M\to\infty}\Bigg\langle \sum_{\bsigma}\frac{\delta\left(\mathbb{H}\ast \bsigma = \mathbf{1}\right)\prod^N_{i=1}p_{b}(\sigma_i)\prod^M_{i=N+1}p_{\frac{1}{2}}(\sigma_i)}{\sum_{\bsigma}\delta\left(\mathbb{H}\ast \bsigma = \mathbf{1}\right)\prod^N_{i=1}p_{b}(\sigma_i)\prod^M_{i=N+1}p_{\frac{1}{2}}(\sigma_i)}f\left(\left\{\sigma_i\right\}\right)\Bigg\rangle_{\mathbb{H}}\:,
\end{eqnarray}
with
\begin{eqnarray}
-\beta f(\left\{\sigma_i\right\}) &=& \frac{1}{M}
\int\prod^M_{i=1}dh_i\prod^N_{i=1}p_{b}(h_i|\sigma_i)\prod^M_{i=N+1}p_{\frac{1}{2}}(h_i|\sigma_i)\log \mathcal{Z}\left(\left\{h_i\right\},\mathbb{H}\right) \:.
\end{eqnarray}
with the partition function
$\mathcal{Z}\left(\left\{h_i\right\},\mathbb{H}\right)$ given by
(\ref{eq:PartSum}). This expression, describing an average over
parity check matrices and input codewords, can be dealt with using
the replica method \cite{par1987}. We replicate
the $\sigma$-variables $g$ times to $\bsigma =
\left(\sigma^1,\cdots,\sigma^g\right)$ and the $\nu$ variables $n$
times to $\bnu = \left(\nu^1,\cdots,\nu^n\right)$. The free energy
per bit (\ref{eq:freeQ}) is then given by an extremization problem:
\begin{eqnarray}
 -\beta \overline{f} &=&\lim_{n\rightarrow 0} \frac1n{\rm extr}_{P,\hat{P}}\Psi\Big\{P(\bnu, \bsigma), \hat{P}(\bnu, \bsigma)\Big\}.
\end{eqnarray}
The function $\Psi$ equals
\begin{eqnarray}
  \fl \Psi\left\{P(\bnu,\bsigma),\hat{P}(\bnu, \bsigma) \right\} = -C\sum_{\bnu, \bsigma}\hat{P}(\bnu,\bsigma)P(\bnu,\bsigma)  + C -\frac{C}{K}
 \nonumber \\
 + \frac{C}{K}\sum_{\bnu_1,\bsigma_1,\bnu_2, \bsigma_2,\cdots,\bnu_K, \bsigma_K}\prod^K_{l=1}P(\bnu_l, \bsigma_l)\prod^n_{\alpha=1}\delta\left(\prod^K_{l=1}\nu^{\alpha}_l, 1\right) \prod^g_{\zeta =1}\delta\left(\prod^K_{l=1}\sigma^{\zeta}_l, 1\right)
\nonumber \\
+ \int da q_b(a) \log\left\{\sum_{\bnu,\bsigma}\prod^g_{\zeta=1}p_b(\sigma^{\zeta})\Big\langle \left(\hat{P}(\bnu, \bsigma)\right)^C  \exp\left[\beta h \sum_{\alpha}\nu^{\alpha}\right]\Big\rangle_{h|\sigma^1,a}   \right\}\:,
\label{eq:FreeE}
\end{eqnarray}
with
\begin{eqnarray}
q_b(a) &=& \left(1-\frac{C}{K}\right)\delta(a-b) + \frac{C}{K}\delta(a-\frac{1}{2})\:.
\end{eqnarray}
The order parameters $P\left(\bnu, \bsigma\right)$ and
$\hat{P}\left(\bnu,\bsigma\right)$ are solutions of the self-consistent
equations
\begin{eqnarray}
\fl \hat{P}(\bnu, \bsigma) = \sum_{\bnu_1, \bsigma_1,\cdots, \bnu_{K-1}, \bsigma_{K-1}}\prod^{K-1}_{l=1}P(\bnu_l, \bsigma_l)
\prod^n_{\alpha=1}\delta\left(\nu^{\alpha}
\prod^{K-1}_{l=1}\nu^{\alpha}_l, 1\right) \prod^g_{\zeta= 1}\delta\left(\sigma^{\zeta}\prod^{K-1}_{l=1}\sigma^{\zeta}_l , 1\right)\:,\nonumber \\ \label{eq:selfcH}
\\
\fl P(\bnu, \bsigma) =\int da \ 
q_b(a)\frac{\prod^g_{\zeta=1}p_a(\sigma^{\zeta}) \Big\langle
\left(\hat{P}(\bnu, \bsigma)\right)^{C-1} \exp\left(\beta h
\sum_{\alpha}\nu^{\alpha}\right)\Big\rangle_{h|\sigma^1,a}
}{\sum_{\bnu,\bsigma}\prod^g_{\zeta=1}p_a(\sigma^{\zeta})
\Big\langle \left(\hat{P}(\bnu, \bsigma)\right)^{C} \exp\left(\beta h
\sum_{\alpha}\nu^{\alpha}\right)\Big\rangle_{h|\sigma^1,a}} \:.
\label{eq:selfc}
\end{eqnarray}
Inserting (\ref{eq:selfcH}) into (\ref{eq:selfc}) produces a single
self-consistent equation in terms of $P(\bnu,\bsigma)$.
\subsection{Replica Symmetry}

\begin{figure}
\begin{center}
\includegraphics[angle=-90, width=.6 \textwidth]{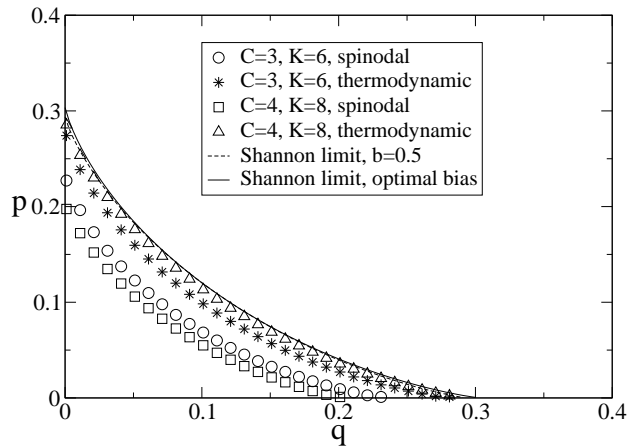}
\caption{Critical noise level lines in the $(p,q)$-parameter space
for symbol-wise MAP decoding and an unbiased source.
Two Gallager codes of rate $R=1/2$ are compared. The channel capacity
obtained from maximizing the mutual information over the input bias
$b$ versus the mutual information for an unbiased source ($b=1/2$) are indistinguishable.}
\label{fig:spinNish}
\end{center}
\end{figure}

For the joint distribution of the replicated spin variables
$P(\bnu,\bsigma)$ we now write
\begin{equation}
P(\bnu,\bsigma) = \int da\ q_b(a)\ P(\bsigma|a)\ P(\bnu|\bsigma,a) \:. \label{eq:Pbnubsigma}
\end{equation}
In order to take the limit $n\rightarrow 0$, one has to make an
assumption for the form of the distribution $P(\bnu|\bsigma,a)$. The
simplest such ansatz corresponds to replica symmetry, i.e.\@
assuming that the $\alpha$-replica indices with respect to the
noise variables are inter-changable. More concretely we write
\begin{eqnarray}
 P(\bnu|\bsigma,a) = \int dx\, \pi(x|\bsigma,a) \prod_{\alpha=1}^n
 \mathcal{Q}(\nu_\alpha|x)\:,\nonumber \\
\mathcal{Q}(\nu|x) = \frac{\exp\left(\beta x\nu\right)}{2\cosh(\beta x)}\:, \label{eq:SigmaAnsatz}
\end{eqnarray}
for some $\pi(x|\bsigma,a)$ with $\int dx\,\pi(x|\bsigma,a)=1$. Using this ansatz we can
convert the self-consistent equation of $P(\bnu,\bsigma)$ into one for the
density $\pi(x|\bsigma,a)$ and one for $P(\bsigma|a)$, namely
\begin{eqnarray}
\fl P(\bsigma|a) =\frac{\prod_{\zeta}p_{a}(\sigma^{\zeta})\left(\int \prod_{l}da_l q_b(a_l)\sum_{\bsigma_1, \bsigma_2, \cdots, \bsigma_{K-1}}\delta\left(\bsigma\prod_{l}\bsigma_l;1\right)\prod^{K-1}_{l=1}P(\bsigma_l|a_l) \right)^{C-1}}{\sum_{\bsigma}\prod_{\zeta}p_{a}(\sigma^{\zeta})\left(\int \prod_{l}da_lq_b(a_l)\sum_{\bsigma_1, \bsigma_2, \cdots, \bsigma_{K-1}}\delta\left(\bsigma\prod_{l}\bsigma_l;1\right)\prod^{K-1}_{l=1}P(\bsigma_l|a_l) \right)^{C}} \:,\nonumber \\
 \label{eq:PSelfc} \\
 \fl \pi(x|\bsigma, a) =   \prod^{C-1}_{r=1}\frac{\int \prod_{l}da^r_lq_b(a^r_l) \sum_{\bsigma^r_1, \bsigma^r_2, \cdots, \bsigma^r_{K-1}}\delta\left(\bsigma\prod_{l}\bsigma^r_l;1\right)\prod_{l}P(\bsigma^r_l|a^r_l)}{\int \prod_{l}da_lq_b(a_l)\left(\sum_{\bsigma'_1, \bsigma'_2, \cdots, \bsigma'_{K-1}}\delta\left(\bsigma'\prod_{l}\bsigma'_l;1\right)\prod_{l}P(\bsigma'_l|a_l)\right)}
\nonumber \\
 \times
 \int \prod^{C-1}_{r=1}\prod^{K-1}_{l=1}dx^r_l\pi(x^r_l|\bsigma^r_l, a^r_l)\int dh p(h|\sigma^1,a)\delta\left[x - u_{\beta}\Big(\{x_l^r\},h\Big) \right] \:.
\nonumber \label{eq:piBsigma}
\end{eqnarray}
We defined the messages
\begin{equation}
u_{\beta}\Big(\{x_l^r\},h\Big) = h + \frac{1}{\beta}\sum^{C-1}_{r=1}\atanh\left(\prod^{K-1}_{l=1}\tanh(\beta x^r_l)\right)\:.
\label{eq:belief}
\end{equation}
To take the limit $g\rightarrow 0$ we make the following assumptions on the $\bsigma$-dependencies
\begin{eqnarray}
\fl P(\bsigma|a) &= \int dy\ \eta(y|a)\ \prod^g_{\zeta=1}\mathcal{Q}\left(\sigma^{\zeta}|y\right)  \:,\label{eq:Pansatz} \\
\fl \pi(x|\bsigma,a) &=  \int dz\ P(z|\bsigma,a)\ \pi(x|\sigma^1,z,a)
\nonumber \\
&= \frac{1}{P(\bsigma|a)}\int dz\  \theta(z|a)\ P(\bsigma|z,a)\ \pi(x|\sigma^1,z,a)
\nonumber \\
&= \left[\int dy\ \eta(y|a)\ \prod^g_{\zeta=1}\mathcal{Q}\left(\sigma^{\zeta}|y\right)\right]^{-1}
\int dz\  \theta(z|a)\ \pi(x|\sigma^1,z,a) \prod^g_{\zeta=1}\mathcal{Q}\left(\sigma^{\zeta}|z\right) \:,
 \nonumber \\\label{eq:piAnsatz}
\end{eqnarray}
with $\int dz \theta(z|a) = 1$ and $\int dx  \pi(x|\sigma,a,z) = 1$.
The distribution $\eta(y|a)$ fullfills the self-consistent equation
\begin{eqnarray}
\fl \eta(y|a) =\int \prod^{C-1}_{r=1}\prod^{K-1}_{l=1} da^r_l\ q_b(a^r_l) \prod^{C-1}_{r=1}\prod^{K-1}_{l=1} dy^r_l\ \eta(y^r_l|a^r_l)\  \delta\left(y - u_1\left(\left\{y^r_l\right\},y_0(a)\right)\right) \:,\label{eq:eta}
\end{eqnarray}
with $y_0(b) = \frac{1}{2}\log\left(\frac{b}{1-b}\right)$.  The
distribution  $\theta(z|a)$  turns out to be also a solution
of the equation (\ref{eq:eta}). The distributions
$\pi(x|\sigma,z,a)$ are given through the equations
\begin{eqnarray}
 \fl\pi(x|\sigma,z,a) =  \frac{\int\left(\prod_{r,l}\mathcal{D}_b\,a^r_l  \:\mathcal{D}_{\theta, a^r_l}\, z^r_l \right)\delta\left(z- u_1\left(\left\{z^r_l\right\},y_0(a)\right)\right) \sum_{\left\{\sigma^r_l\right\}}\prod_{r}P(\left\{\sigma^r_l\right\}|\sigma,\left\{z^r_l\right\})}{\int\left(\prod_{r,l}\mathcal{D}_b\,a^r_l  \:\mathcal{D}_{\theta, a^r_l}\, z^r_l \right)\delta\left(z- u_1\left(\left\{z^r_l\right\},y_0(a)\right)\right)}
\nonumber \\
\times
\int \prod_{r,l}dx^r_l\pi(x^r_l|\sigma^r_l,z^r_l,a^r_l)\int dh p(h|\sigma,a)\delta\left(x-u_{\beta}\left(\left\{x^r_l\right\},h\right) \right)\:, \label{eq:selfcDistriFinal}
\end{eqnarray}
with $da\: q_b(a) = \mathcal{D}_b\, a$ and $dz\: \theta(z|a) = \mathcal{D}_{\theta,a}\, z$ and
\begin{eqnarray}
 P(\left\{\sigma_l\right\}|\sigma,\left\{z_l\right\}) &=& \frac{\delta\left(\sigma\prod_{l}\sigma_l;1\right)\prod_l\mathcal{Q}(\sigma_l|z_l)}{\sum_{\sigma_1,\cdots,\sigma_{K-1}}\delta\left(\sigma\prod_{l}\sigma_l;1\right)\prod_l\mathcal{Q}(\sigma_l|z_l)}\:.
\end{eqnarray}
Equations (\ref{eq:eta}) and (\ref{eq:selfcDistriFinal}) are the main equations from which the various thermodynamic quantities in this section will be derived.
We remark that the distribution $\pi(x|\sigma,z,a)$ gives the
distribution of cavity fields given the value of $z$ on the corresponding link and the value of $\sigma$ and $a$ on the corresponding site. Finally we have to average over $\sigma$ and $z$.  Substitution of the ans\"atze (\ref{eq:Pbnubsigma}), (\ref{eq:SigmaAnsatz}), (\ref{eq:Pansatz}) and (\ref{eq:piAnsatz}) in the expression (\ref{eq:FreeE}) of the free energy leads to, after taking the limits $n\rightarrow 0$ and  $g\rightarrow 0$
\begin{eqnarray}
 \fl-f_{RS} =\left(\frac{C}{K}\left(K-1\right)\right) \mathbb{E}^{(K)}_{b}\left[\Delta F^{(K)}_{RS}\left(\left\{x_l\right\}\right)\right] - \mathbb{E}^{(1)}_b\left[\Delta F^{(1)}_{RS}\left(\left\{x^r_l\right\}; h\right)\right]\:, \label{eq:finalFreeE}
\end{eqnarray}
with
\begin{eqnarray}
\fl \mathbb{E}^{(K)}_{b} \left[g\left(\left\{x_l\right\}\right)\right] = \int \prod^K_{l=1}\mathcal{D}_b\, a_l \: \mathcal{D}_{\theta, a_l}\,z_l\sum_{\sigma_1,\cdots, \sigma_{K}}P\left(\left\{\sigma_l\right\}|\left\{z_l\right\}\right)
\prod^K_{l=1}dx_l\pi(x_l|\sigma_l,z_l,a_l) g\left(\left\{x_l\right\}\right) \:, \nonumber \\
\\
\fl \mathbb{E}^{(1)}_b\left[f(\left\{x^r_l\right\}; h)\right] = \int \mathcal{D}_b(a) \left(\prod_{r,l}\mathcal{D}_b\,a^r_l  \:\mathcal{D}_{\theta, a^r_l}\, z^r_l \right)
\sum_{\sigma} p_a(\sigma)\prod^{C}_{r=1}\sum_{\sigma^r_1\cdots\sigma^r_{K-1}}P\left(\left\{\sigma^r_l\right\}|\sigma, \left\{z^r_l\right\}\right)
\nonumber \\
 \times
 \int \left(\prod_{r,l}dx^r_l \pi(x^r_l|\sigma^r_l,z^r_l,a^r_l)\right)\int dh p(h|\sigma,a)f(\left\{x^r_l\right\}; h)\:,
\end{eqnarray}
and where we used the abbreviations
\begin{eqnarray}
  P(\left\{\sigma_l\right\}|\left\{z_l\right\}) &=& \frac{\delta\left(\prod_{l}\sigma_l;1\right)\prod_l\mathcal{Q}\left(z^r_l|\sigma_l\right)}{\sum\delta\left(\prod_{l}\sigma_l;1\right)\prod_l\mathcal{Q}\left(z^r_l|\sigma_l\right)} \:,\\
\Delta F^{(K)}_{RS} &=& -\frac{1}{\beta}\log\left(1+\prod^K_{l=1}\tanh \beta x_l\right)  + \frac{1}{\beta}\log(2)  \:,
\\
\Delta F^{(1)}_{RS} &=& -\frac{1}{\beta}\log \left(\sum_{\tau} e^{\beta h \tau}\prod^C_{r=1}\frac{1}{2}\left( 1 + \tau \prod^{K-1}_{l=1}\tanh \beta x^r_l\right)\right) \:.
\end{eqnarray}
In the unbiased case $b = \frac{1}{2}$, we have the stable solution $\theta(z|a) = \delta(z)$ and $\pi(x|\sigma,z,a) = \pi(x|\sigma)$ with
 \begin{eqnarray}
 \fl\pi(x|\sigma) = \prod_{r=1}^{C-1} \left(\sum_{\sigma^r_1,\cdots, \sigma^r_{K-1}}\frac{\delta\left(\sigma\prod_{l=1}^{K-1}\sigma^r_l\right)}{2^{K-2}} \int \prod^{K-1}_{l=1}\pi(x^r_l|\sigma^r_l)\right)
\nonumber \\
\times
\int dh p(h|\sigma,\frac{1}{2}) \delta(x-u_{\beta}\left(\left\{x^r_l\right\},h\right))\:,  \label{eq:distriBACdistri}
 \end{eqnarray}
and
\begin{eqnarray}
 \fl -f_{RS} =\left(\frac{C}{K}\left(K-1\right)\right) \mathbb{E}^{(K)}\left[\Delta F^{(K)}_{RS}\left(\left\{x_l\right\}\right)\right] - \mathbb{E}^{(1)}\left[\Delta F^{(1)}_{RS}\left(\left\{x^r_l\right\}; h\right)\right]\:,\label{eq:RSFreeEnergy}
\end{eqnarray}
with
\begin{eqnarray}
 \fl \mathbb{E}^{(K)}_{RS} \left[g\left(\left\{x_l\right\}\right)\right] = \left(\sum_{\sigma_1,\cdots, \sigma_{K}}\frac{\delta\left(\prod_{l}\sigma_l\right)}{2^{K-1}} \int \prod^{K}_{l=1}\pi(x_l|\sigma_l)\right) g\left(\left\{x_l\right\}\right)\:,\\
\fl \mathbb{E}^{(1)}_{RS}\left[g(\left\{x^r_l\right\}; h)\right] =  \sum_{\sigma} \left(\prod^{C}_{r=1}\sum_{\sigma^r_1\cdots\sigma^r_{K-1}}\frac{\delta\left(\sigma\prod_{l}\sigma^r_l;1\right)}{2^{K-2}}\right)
\nonumber \\
\times \int  \prod_{r,l} dx^r_l \pi(x^r_l|\sigma^r_l)\int dh p(h|\sigma,\frac{1}{2})\: g(\left\{x^r_l\right\}; h) \:.
\end{eqnarray}

\begin{figure}
\begin{center}
\includegraphics[angle=-90, width=.6 \textwidth]{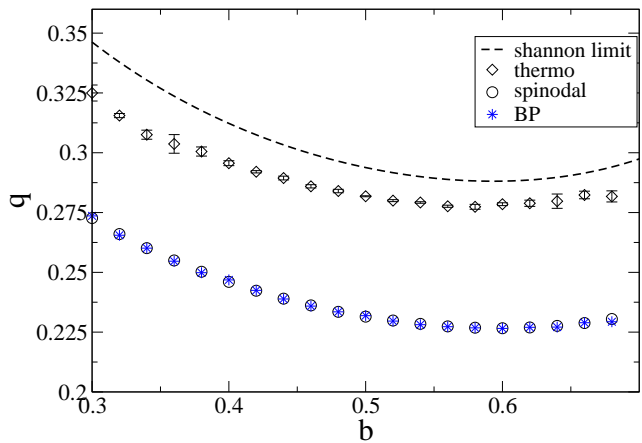}
\caption{Thresholds of the noise variable $q$ as a function of the bias for
Z-channels with an encoding strategy $(C,K)=(3,6)$. The thermodynamic and spinodal lines obtained through population dynamics are compared with the Shannon limit and the results obtained by belief propagation.  The
location of the minimum lies at $b>\frac12$.  The spinodal and thermodynamic lines
are calculated for a set of 100 $\pi$-populations each of 1000 x-fields.  The BP-points are calculated on one graph instance with $10^6$ sites.}
\label{fig:spinNishB}
\end{center}
\end{figure}
We see that for unbiased sources the formulas (\ref{eq:selfcDistriFinal}) and (\ref{eq:finalFreeE}) are much simplified to (\ref{eq:distriBACdistri}) and (\ref{eq:RSFreeEnergy}).  Generalization of these formulas to irregular graphs is straightforward.
We note that $\Delta F^{(K)}_{RS}$ and $\Delta
F^{(1)}_{RS}$ correspond, in the framework of the cavity
method \cite{meZ2003}, to the free energy shifts due to link- and
site-addition respectively.   Equation (\ref{eq:distriBACdistri})
is also known as the `density evolution' equation (while the
equivalent (\ref{eq:cav2}) of \ref{sec:cav} which refers to
a single graph instance is termed as the `belief propagation'
equation). 

We see that the state $\pi(x|\sigma, z, a) = \delta(x-\infty)$ is always a solution to
(\ref{eq:selfcDistriFinal}) and gives $\rho=1$. If the initial state lies in the basin of
attraction of this solution, errorless decoding is possible. We will term this the
ferromagnetic solution. The ferromagnetic state has a free energy equal to
\begin{eqnarray}
 f_{ferro} = \epsilon_{ferro} &=& -\int da q_b(a) \int \prod_{r,l}da^r_lq_b(a^r_l) \int  \prod_{r,l}dz^r_l \theta(z^r_l|a^r_l)
\nonumber \\
&& \times   \sum_{\sigma} p_a(\sigma)\prod^{C}_{r=1}\sum_{\sigma^r_1\cdots\sigma^r_{K-1}}P\left(\left\{\sigma^r_l\right\}|\sigma, \left\{z^r_l\right\}\right) \langle h \rangle_{h|\sigma,a} \:. \label{eq:epsilonFerro}
\end{eqnarray}
From equations (\ref{eq:RSFreeEnergy}), (\ref{eq:epsilonFerro}) we
see that for $\kappa = 0$, the free energy becomes $-\infty$. To
avoid these infinities we will solve the problem for $\kappa
\approx 0$ and look at quantities that are finite for
$\kappa\rightarrow 0$.

\begin{table}
\begin{tabular}{cc|cccccc}
\hline \\
& &      \multicolumn{2}{c}{$\kappa = 1$}&  \multicolumn{2}{c}{$\kappa = 0$}& \multicolumn{2}{c}{$\kappa = 0.1$}\\
 $C$& $K$&  $q_d$&$q_c$ & $q_d$&$q_c$&  $q_d$&$q_c$\\
 \hline \\
5 & 6 &  0.13739(5)& 0.26436(3)&0.35546(2)& 0.70400(5)& 0.28709(2)& 0.54800(1)\\
3 & 4 &  0.16703(1) &0.20959(1) & 0.45580(2) &0.57591(4)& 0.35426(1) &0.44167(1) \\
4 & 6 & 0.11692(1) &0.17245(1) &0.30802(2) & 0.47132(5)& 0.24615(1)& 0.36373(2) \\
3 & 6  & 0.08406(1) & 0.09972(1) &0.23146(2)&0.27880(1) & 0.17977(2) &0.21329(2) \\
4 & 8 &0.07681(1) &0.10717(1)&0.20056(2)&0.2905(1) & 0.16137(2)&0.22635(2) \\
\hline
\end{tabular}
\caption{The spinodal ($q_d$) and thermodynamic ($q_c$) critical noise levels calculated within the replica symmetric
ansatz at $T = 1$ and with an unbiased source.  The variable $\kappa
= p/q$ controls the amount of symmetry in the channel noise.  The thresholds for the Z-channel are calculated with
$\kappa\sim\mathcal{O}(10^{-8})$.} \label{table:points}
\end{table}

\begin{figure}[h!]
\begin{center}
\includegraphics[angle=-90, width=.6 \textwidth]{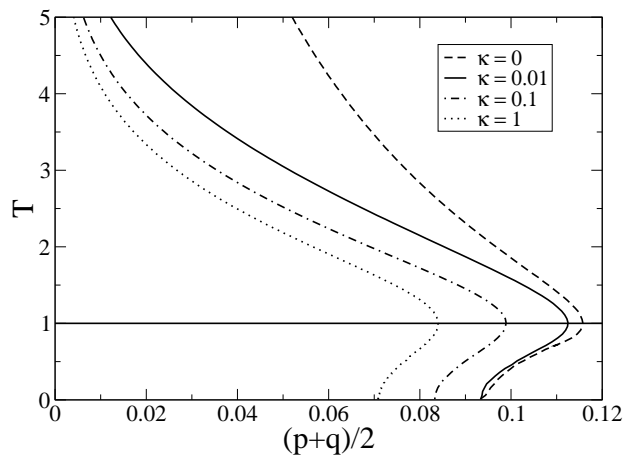}
\caption{The $(T,(p+q)/2)$-phase diagram for the spinodal transition lines.  At these lines the $\pi(x|\sigma) = \delta(x)$ state lies at the boundary of the ferromagnetic basin of attraction.  The lines are
calculated within the replica symmetric approximation for a $(C,K)=(3,6)$ regular
Gallager code with an unbiased source. The variable $\kappa=p/q$ controls
the amount of asymmetry in the channel noise. The Nishimori line, $T=1$, is visualized.} \label{fig:RSphase}
\end{center}
\end{figure}

We solve the coupled set of equations (\ref{eq:eta}) and
(\ref{eq:selfcDistriFinal}) using a methodology similar qto
`population dynamics' \cite{Mez2001}. We first derive the stationary
distribution for the density (\ref{eq:eta}) describing a population
of $y$-fields $\{y_1,\ldots,y_n\}$. We remark that since
$\theta(z|a)=\eta(z|a)$ we need not to update separately a
population of $z$-fields. For every $y$-field $y_i$ we associate a
$\pi^{(i)}$-population of $x$-fields, namely
$y_i\to\{x_1^{(i)},\ldots,x_m^{(i)}\}$. A stationary solution for
the population of populations of $x$-fields is found through
the following algorithm:
\begin{enumerate}
 \item We select $(C-1)(K-1)$ fields from the
 $y$-population: $\{y_\ell\}$ with $\ell\in\mathcal{S}$ and $\mathcal{S}$ the set of chosen indices $\mathcal{S} =\{i_1,\ldots,i_{(C-1)(K-1)}\}$.
 \item We calculate a new $z$-field according to its update
 rule: $z_{\star}=u_1\Big(\{y_\ell\}_{\ell\in\mathcal{S}},y_0(a)\Big)$
 \item We use the $(C-1)(K-1)$ populations of $x$-fields indexed by $\mathcal{S}$ to calculate a new population of $x$-fields
 with the update rule $x_\star =  u_\beta\Big(\{x_\ell\}_{\ell\in\mathcal{S}},h\Big)$
\item We select at random an index $j\in[0,n]$ and replace the
$j$-th member of the $y$-population by $z_\star$ and the $j$-th
member of the $x$-population by $x_{\star}$.
\end{enumerate}

We start with the initial distribution $\pi(x|\sigma,z,a) =
\delta(x)$. This corresponds to a state with no a priory knowledge
on the message $\bsigma^0$.  At low temperature this distribution
converges to the ferromagnetic state.  Increasing the noise at a
constant temperature we find that at some critical noise level
$(p_d, q_d)$ a second solution appears (suboptimal solution) with
$\rho<1$.  We remark that below the $(p_d,q_d)$-threshold, the
ferromagnetic state is the only stable state for all initial
conditions.  From an algorithmic point of view $(p_d,q_d)$ is
the threshold to successful decoding with the belief propagation
algorithm.  The thermodynamic transition is
determined by the point $(p_c, q_c)$ where the free energy of the
suboptimal solution becomes lower then the free energy of the
ferromagnetic solution.  At $T=1$ this determines the limit for
maximum likelihood decoding.  In figure \ref{fig:spinNish} we see
that when we increase $C$ and keep the rate constant, the limit for
maximum likelihood decoding increases but the limit for belief
propagation decoding decreases.  Indeed, we found that in the
RCM the paramagnetic state is always stable.  We find that, in contrary
to the cases of symmetric channels, the bias $b$ in the input
signal influences the decoding process. However, comparing the
critical noise levels $(p_s, q_s)$, above which errorless decoding is impossible for all type of encoding processes, between an optimal biased source and an unbiased source,  see figure \ref{fig:shannonLimB}, we see that bias has few influence.  In figure
\ref{fig:spinNishB} we show, for a Z-channel, how these thresholds
get influenced by the bias.  For a BSC the minimum lies at
$b=\frac{1}{2}$, whereas for a Z-channel the minimum channel noise
$q$ lies at a point $b>\frac{1}{2}$. We also compared our results obtained through population dynamics with a specific application of the belief propagation algorithm on a specific graph instance.  The results of both methods match very well.  In table
\ref{table:points} we present the spinodal and thermodynamic
critical values for different regular codes and for various degrees
of symmetry. These results are consistent with values found in
\cite{Mont2006,franz} and in good agreement with those of
\cite{Wang2005}.  In figure \ref{fig:RSphase} we present the spinodal transition lines for different values of the inverse
temperature $\beta$ and the parameter $\kappa$.  We find re-entrance effects below the
Nishimori line. We find these re-entrance effects also in the thermodynamic lines.  In figure \ref{fig:entro} we plot the entropy
$s=-\partial f/\partial T$, with $f$ determined through equation
(\ref{eq:RSFreeEnergy}), as a function of the channel parameters. We
see that it becomes negative at the spinodal noise level $(p_d,
q_d)$. The entropy at the Nishimori temperature has a special
meaning as it is the average entropy of the transmitted message once
the received message is known. This is therefore the theoretical
upper limit irrespectively of the decoding dynamics. The energy at
$T=1$, see equation (\ref{eq:energyFerro}), equals the ferromagnetic
energy (\ref{eq:epsilonFerro}).  From this it follows that the
entropy at $T=1$ becomes greater than zero at the critical noise
level $(p_c, q_c)$.  Performing a large $(C,K)$ expansion of (\ref{eq:distriBACdistri}) and (\ref{eq:RSFreeEnergy}), as done for the BSC in \cite{montanari}, we get for the critical noise levels, taking $\kappa = \frac{p}{q}$ constant, 
\begin{eqnarray}
\fl q = q^{(0)}_c + \frac{1}{2\log(2)}\left(1-R\right) \left(\frac{d}{d q^{(0)}} \mathcal{I}(\kappa q^{(0)}_c, q^{(0)}_c)\right)^{-1} 
\nonumber \\ 
\times
\left(\frac{\left(1- q^{(0)}_c\left(\kappa+1\right)\right)^{2}}{\left(1+(\kappa-1)q^{(0)}_c\right)\left(1-(\kappa-1)q^{(0)}_c\right)}\right)^{K} + \mathcal{O}\left(\left(v(q^{(0)}_c)\right)^{K}\right) \label{eq:critNoise}
\end{eqnarray}
with $\mathcal{I}(\kappa q^{(0)}_c, q^{(0)}_c) = R$.  The function $v$ is given by
\begin{eqnarray}
  \fl v(q^{(0)}_c) =\frac{(1-q^{(0)}_c\left(\kappa+1\right))^3}{\left(1+(\kappa-1)q^{(0)}_c\right)\left(1-(\kappa-1)q^{(0)}_c\right)}
\nonumber \\ 
 \times
\max\left\{-2(\kappa-1)q^{(0)}_c, \frac{1-q^{(0)}_c\left(\kappa+1\right)}{(1-(\kappa-1)q^{(0)}_c)(1+(\kappa-1)q^{(0)}_c)}\right\}
\end{eqnarray}
From (\ref{eq:critNoise}) we find for a (3,6)-code when $\kappa=1$, $p_c = 0.103968$ and $\kappa=0$, $p_c = 0.284897$.  For a (3,4)-code we find when $\kappa=1$, $p_c= 0.213414$ and when $\kappa=0$, $p_c = 0.579815$.

\subsection{The entropy crisis}

\begin{figure}
\begin{center}
\includegraphics[angle=-90, width=0.6 \textwidth]{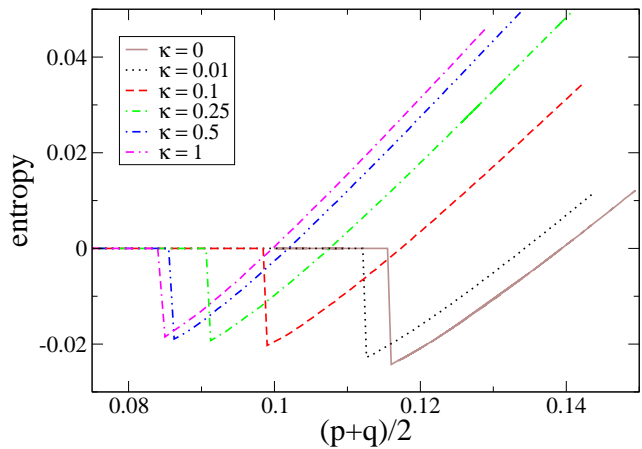}
\caption{The entropy $s$ as a function $(p+q)/2$ of the stationary
solution of the density evolution equation (\ref{eq:distriBACdistri}), starting from $\pi(x|\sigma) = \delta(x)$.  The lines are calculated at the Nishimori temperature for a (3,6)-regular code. From
left to right: $\kappa = \{0,0.01,0.1,0.25,\frac{1}{2},1\}$. 
The sub-optimal ferromagnetic solution at a certain
critical noise level emerges with a negative entropy. This solution
becomes the thermodynamic one when the entropy becomes positive.}\label{fig:entro}
\end{center}
\end{figure}

From figure \ref{fig:entro} we have learned that the entropy can
indeed become negative for the asymmetric channel.  We also found re-entrance effects in the thermodynamic transition lines.  This indicates that something is missing in our solution.  In the SK-model
\cite{par1987},  the negative entropy in the ground state is an
indication that the replica symmetric formalism is incorrect.  In
general $p$-spin models, see \cite{Franz2001}, we have at a
temperature $T_d$ a transition from a paramagnetic phase to a one
step replica symmetry breaking phase and at a temperature $T_K<T_d$
an entropy crisis corresponding with the vanishing of the
configurational entropy.  Here, though, because of the
infinitely strong interactions, the first phase transition
will not appear \cite{Martin2004}.  We will have an entropy
crisis just like it occurs in the RCM.  Because we are interested in
the typical behavior of the system we should define the typical free
energy $f_t(\beta)$, as
\begin{equation}
 f_t(\beta) = \left\{ \begin{array}{ccc} \overline{f}_{RS}(\beta) &\rm if& s_{RS}\geq 0\\
 \overline{f}_{RS}(\beta_f) &\rm if& s_{RS}< 0\end{array}\right. \:, \label{eq:frozenf}
\end{equation}
where $\beta_f$ is the inverse temperature at which the system
freezes in the lowest energy paramagnetic configuration, i.e.\@
$s(\beta_f) = 0$.
\begin{figure}[h]
\begin{center}
\includegraphics[angle=-90, width=.6 \textwidth]{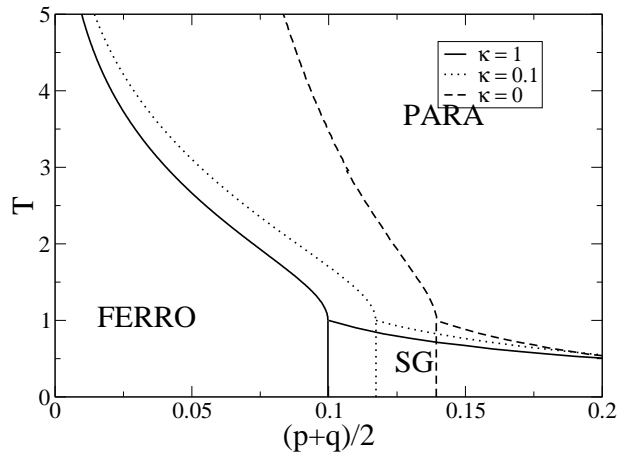}
\caption{The  thermodynamic (T,(p+q)/2)-phase diagram in the frozen ansatz for a regular $(C,K)=(3,6)$ code and unbiased
source.  A FERRO, PARA and SG phase occur.}\label{RSThermophase}
\end{center}
\end{figure}
Because of the hard constraints, we have indeed that following frozen ansatz, see \cite{montanari},
 \begin{eqnarray}
 \fl P(\bnu|\bsigma,a) =\sum_{\left\{n^{\gamma}\right\}}\left(\int dx \pi_m(x|\bsigma,a)
\prod^{n/m}_{\gamma=1}\mathcal{Q}(n^{\gamma}|x)\right)
\left(\prod^{n/m}_{\gamma=1}\prod^{m}_{\alpha=1}\delta\left[\nu_{\gamma,
\alpha}, n^{\gamma}\right]\right)\:,\label{eq:frozenansatz}
\end{eqnarray}
with $m\in [0,1]$, fullfills the selfconsitent equations (\ref{eq:selfc}) . Using this ansatz in  the self-consistent
equations (\ref{eq:selfc}) and the free energy expression (\ref{eq:FreeE}), we
find back the replica symmetric equations (\ref{eq:selfcDistriFinal}) and
 (\ref{eq:finalFreeE}) with $\beta \rightarrow \beta m$.  The
extremization condition $\frac{\partial f_{RS}}{\partial m}=0$
corresponds to the zero entropy condition, which for $m\in[0,1]$ can
only be fullfilled when $\beta\geq\beta_f$.  When $\beta<\beta_f$ we
have $m=1$, because there the free energy is indeed maximal.  This corresponds with the frozen scenario of (\ref{eq:frozenf}).
We will call the phase where the entropy
is zero and $\rho<1$ the spin glass phase and the  phase where
$s>0$ and $\rho<1$ the paramagnetic phase.  In the spin glass phase the thermodynamic average is dominated by a subexponential  amount (in the system size) of codewords whereas in the paramagnetic phase the average is dominated by an exponential amount of codewords.
In figure \ref{RSThermophase} we plot the full thermodynamic phase diagram of the
system in the space of $(T,\frac12(p+q))$ for a regular
$(C,K)=(3,6)$ code with an unbiased source and three different levels of symmetry in
the channel noise.   The re-entrance effects have disappeared because of the frozen ansatz.
\section{Non-convergence regions of belief propagation and endogeny}
\label{sec:bp}

\begin{figure}
\begin{minipage}{.45\textwidth}
\begin{center}
\includegraphics[angle=-90, width = 1\textwidth]{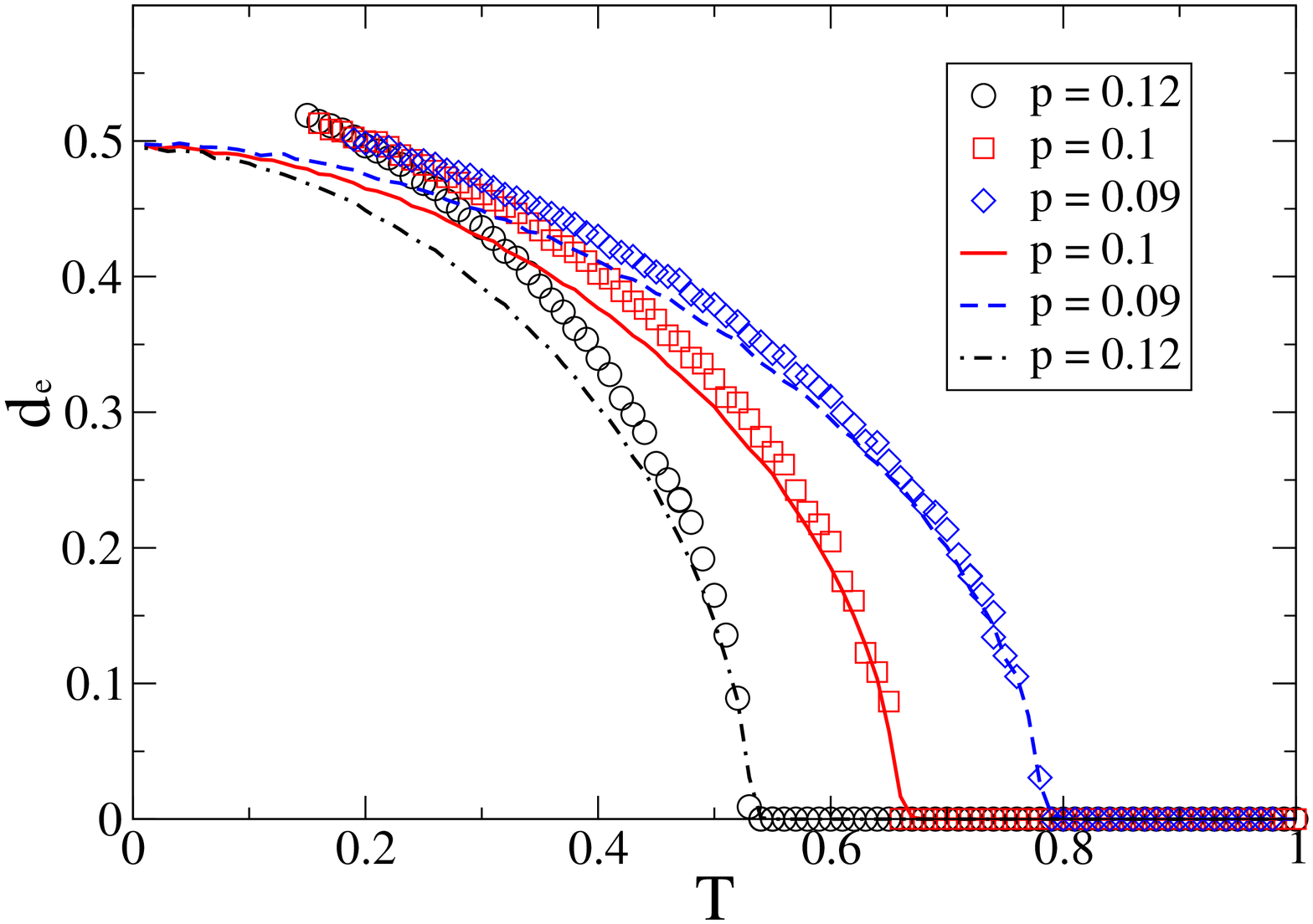}
\end{center}
\end{minipage}
\hfill
\begin{minipage}{.45\textwidth}
\begin{center}
\includegraphics[angle=-90,width= 1\textwidth]{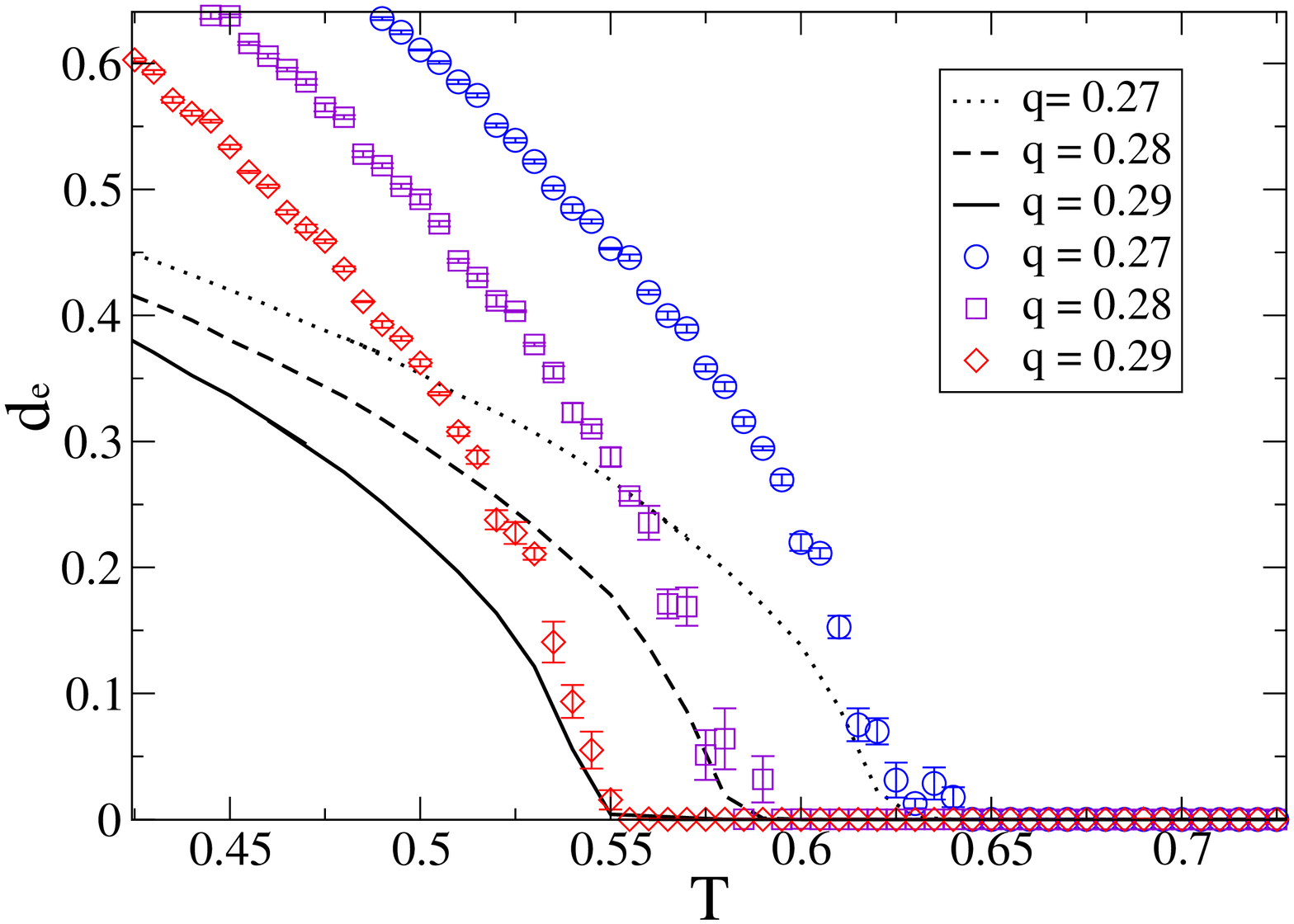}
\end{center}
\end{minipage}
\caption{The endogeny parameter $d_e$ as a function of
the temperature $T$ for a regular $(C,K)=(3,6)$ Gallager code. Left: binary symmetric channel, right: Z-channel. Lines
represent the endogeny parameter calculated
through population dynamics (the $d_e$ presented for the Z-channel is for $\sigma=-1$). For $T$ such that $d_e>0$ no
meaningful solution exists to the decoding equations. Markers
indicate the values $d_e$ calculated for the log-likelihood ratios of the belief propagation fields at different times steps.}
\label{fig:endo1}
\end{figure}

Although the freezing scenario presented above seems to
explain the thermodynamical phase diagram completely, the dynamics
of the system can be disturbed by a clustering of the phase space.
To investigate this more closely we will consider a system
composed of two copies of the original dynamic variables. These are
embedded on the same graph and interact with the same quenched fields. In this setting, convergence of the recursive decoding
equations can be quantified through the resulting statistics of the
joint system.

We remark that the convergence of the belief propagation
equations can be seen as one example of a general class of problems
in which one is interested in the stationary density that solves a
distributional fixed-point problem. This problem and its
applications to various fields has been studied in a rigorous way by
Aldous and Bandyopadhyay \cite{aldous2005} from the viewpoint of
theoretical statistics.  This link between the two fields has been observed in \cite{RivThesis}.

\begin{figure}
\begin{center}
\includegraphics[angle=-90, width=.6 \textwidth]{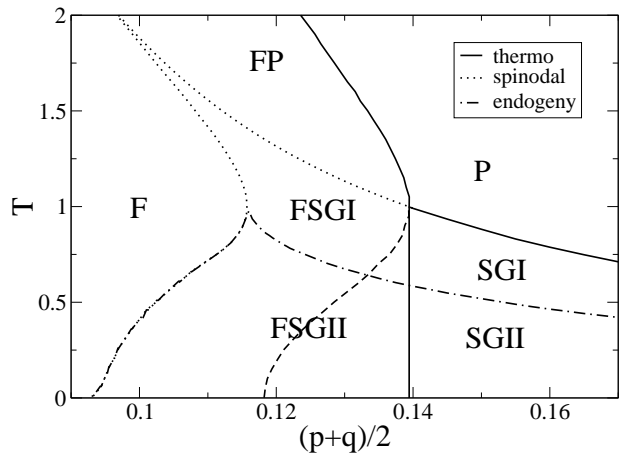}
\caption{Full $(T, (p+q)/2)$-phase diagram for a (3,6)-encoding scheme over a
Z-channel.  The solid lines indicate thermodynamic phase transitions and
the dotted lines represent spinodal transitions.   The dashed line
determines the thermodynamic transition in the replica symmetric approximation. The
vertical line determines the same transition in the frozen ansatz.
The dashed-dotted endogeny line bounds the region below which the BP algorithm stops
converging. }\label{fig:kappaFull0}
\end{center}
\end{figure}

\begin{figure}
\begin{center}
\includegraphics[angle=-90, width=.6 \textwidth]{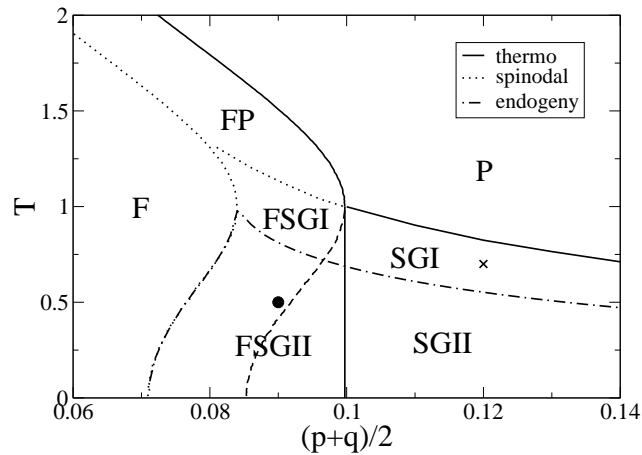}
\caption{The same lines as presented in figure
\ref{fig:kappaFull0} for a BSC and a (3,6)-encoding scheme.
Two additional points are added where a
full 1RSB calculation is performed.} \label{fig:kappaFull}
\end{center}
\end{figure}

In the case of a two-replica system we define a partition
function of a form similar to (\ref{eq:PartSum}): 
\begin{eqnarray}
\fl Z\left(\left\{h_i\right\},\mathbb{H}\right) = \sum_{\bnu,\bmu}\exp\left(\gamma\sum_{\langle i_1, i_2, \cdots, i_K\rangle}\mathcal{T}_{i_1, i_2, \cdots, i_K}\left(\nu_{i_1}\nu_{i_2}\cdots\nu_{i_K} +\mu_{i_1}\mu_{i_2}\cdots\mu_{i_K} \right)\right)
\nonumber \\
\times  \exp\left(\beta \sum^M_{i=1}\left(h_i\nu_i + h_i\mu_i + \gamma_{\mu\nu}\nu_i\mu_i + \gamma_{\mu}\mu_i  +  \gamma_{\nu}\nu_i\right)\right)   \:,
\end{eqnarray}
with $\gamma\rightarrow \infty$.
Analagously as in section \ref{sec:f} we find
for the unbiased case, the following order parameter equations
\begin{eqnarray}
  \fl \hat{P}\left(\bnu,\bmu|\sigma\right)= \sum_{(\bnu_1,\bmu_1,\sigma_1),\cdots,(\bnu_{K-1},\bmu_{K-1},\sigma_{K-1})}\frac{\delta\left(\sigma\prod^{K-1}_{l=1}\sigma_l;1\right)}{2^{K-2}}\left(\prod^{K-1}_{l=1}P\left(\bnu_l,\bmu_l|\sigma_l\right)\right)
\nonumber \\
 \times \prod_{\alpha}\delta\left(\nu^{\alpha}\prod^{K-1}_{l=1}\nu^{\alpha}_l;1\right)\delta\left(\mu^{\alpha}\prod^{K-1}_{l=1}\mu^{\alpha}_l;1\right) \:,\\
\fl P(\bnu, \bmu|\sigma) = \frac{\Big \langle \left(\hat{P}\left(\bmu, \bnu|\sigma\right)\right)^{C-1} \exp\left(\beta h \sum_{\alpha}\left(\nu^{\alpha} + \mu^{\alpha}\right)\right)\Big\rangle_{h|\sigma}}{\sum_{\sigma}\sum_{\bmu,\bnu}\Big\langle\left(\hat{P}\left(\bmu, \bnu|\sigma\right)\right)^{C} \exp\left(\beta h \sum_{\alpha}\left(\nu^{\alpha} + \mu^{\alpha}\right)\right)\Big\rangle_{h|\sigma}} \label{eq:endogP} \:.
\end{eqnarray}
We introduce the replica symmetric ansatz
\begin{eqnarray}
\fl P(\bnu,\bmu|\sigma) = \int
dx^{(1)}dx^{(2)}\pi(x^{(1)}, x^{(2)}|\sigma) \frac{\exp\left(\beta
x^{(1)}\sum_{\alpha}\nu^{\alpha} + \beta
x^{(2)}\sum_{\alpha}\mu^{\alpha}  \right)}{\left(4\cosh\left(\beta
x^{(1)}\right)\cosh\left(\beta x^{(2)}\right)\right)^n}\:.
\end{eqnarray}
We remark that  a field coupling the $\bmu$ and $\bnu$ variables is not needed because in (\ref{eq:endogP}) the quenched field does not couple $\bmu$ and $\bnu$ variables.  
Substitution of this ansatz in the above self-consistent equations leads to
\begin{eqnarray}
 \fl \pi(x^{(1)},x^{(2)}|\sigma) = \mathbb{E}_{h|\sigma} \int  \prod^{C-1}_{r=1}\left[\sum_{\sigma^r_1,\cdots,\sigma^r_{K-1}} \frac{\delta\left(\sigma\prod^{K-1}_{l=1}\sigma^r_l;1\right)}{2^{K-2}} \prod^{K-1}_{l=1}dx^{(1)}_{r,l}dx^{(2)}_{r,l}\pi(x^{(1)}_{r,l},x^{(2)}_{r,l}|\sigma^r_l)\right]
\nonumber \\
 \times \delta\left[x^{(1)}-u\left(\left\{x^{(1)}_{r,l}\right\},h\right)\right] \delta\left[x^{(2)}-u\left(\left\{x^{(2)}_{r,l}\right\},h\right)\right]\:.
\end{eqnarray}
Now we check whether the distribution $\pi_0(x_1,x_2|\sigma) = \pi_0(x_1|\sigma)\pi_0(x_2|\sigma)$, with $\pi_0(x|\sigma)$ the solution to (\ref{eq:distriBACdistri}), converges to $\pi(x_1,x_2|\sigma) = \pi_0(x_1)\delta\left(x_1-x_2\right)$ (note that all computations are done within RS). We introduce the quantity
\begin{eqnarray}
 d_e(\pi(x_1,x_2|\sigma)) &=& \frac{\int dx_1dx_2 \pi(x_1, x_2|\sigma) |x_1-x_2|}{\int dx_1dx_2 \pi(x_1, x_2|\sigma) \left(\frac{|x_1| + |x_2|}{2}\right)} \:. \label{eq:de}
\end{eqnarray}
We remark that $d_e=0$ corresponds in the two replica formalism to
$q-m^2=0$ and hence corresponds with some sort of spin glass
behavior in the same sense as in the SK model.  This two replica
formalism is, in certain models, proven to be
equivalent with the endogenous property, see \cite{aldous2005}. The
failure of the endogenous property has serious consequences on the
convergence of the BP equations.  We define
\begin{eqnarray}
 \pi^{(t)}_{BP}(x^{(1)}, x^{(2)}) \equiv \frac1M\sum^M_{i=1} \delta\left(x^{(1)}-h^{(t-1)}_{i}\right) \delta\left(x^{(2)}-h^{(t)}_{i}\right)\:,
\end{eqnarray}
with $h^{(t)}_{i}$ the log-likelihood ratio on site $i$ on the $t^{th}$ time step of the BP algorithm (\ref{eq:cav2}). As long as $\lim_{t\rightarrow \infty}d_e(\pi^{(t)}_{BP}(x^{(1)}, x^{(2)}))=0$,  the BP equations converge.  In figure \ref{fig:endo1} we compare the parameter $d_e$ of both formalism and we find that indeed $d_e>0$ when the BP equations stop converging. We call the line marking the
transition from $d_e=0$ to $d_e>0$ the endogeny line.
In figures \ref{fig:kappaFull0} and \ref{fig:kappaFull} we present this line respectively for a Z-channel and a BSC, together with the different thermodynamic and spinodal lines.  In table \ref{table:phases} we give a summary of the various regions of the phase diagrams in figures \ref{fig:kappaFull0} and \ref{fig:kappaFull}.  Performing a high connectivity expansion, like is done in \cite{montanari} we find that the endogeny parameter of the paramagnetic solution is zero.

\begin{table}
\begin{center}
\begin{tabular}{c|ccc}
 &&& \\
& convergence of BP & free energies \\
\hline \\
F  & yes & $\backslash$ \\
FP &  yes& $f_{\rm FERRO}<f_{\rm PARA}$\\
P &  yes& $f_{\rm FERRO}>f_{\rm PARA}$ \\
SGI &  yes& $f_{\rm FERRO}>f_{\rm SG}$ \\
SGII &  no& $f_{\rm FERRO}>f_{\rm SG}$ \\
FSGI & yes& $f_{\rm FERRO}<f_{\rm SG}$ \\
FSGII & no& $f_{\rm FERRO}<f_{\rm SG}$ \\
\end{tabular}
\end{center}
\caption{The different labels used in figures
\ref{fig:kappaFull0} and \ref{fig:kappaFull}.  The convergence of the BP algorithm is indicated.  The free energies of the stable states are compared.}
\label{table:phases}
\end{table}

\section{1RSB ansatz}
\label{sec:1rsb}

From the results of the previous section we know that replica
symmetry fails below a certain temperature. The non-convergence of
the belief propagation equations below a certain temperature reveals
that the amount of solutions of the belief propagation equations (\ref{eq:cav2})
scales exponentially with the system size.  This can be solved using
replica symmetry breaking, which correspond to a more
advanced algorithm. In optimization problems, using insights of 1RSB-effects, practical algorithms have been found (see \cite{SP2005}).  To
count the number of the solutions of (\ref{eq:cav2}), we introduce a Lagrange
parameter $\mu$ conjugate to the free energy of these solutions. We
have a generalized free energy $\Phi$ corresponding with
\begin{eqnarray}
 -\mu\Phi &=& \log\sum_{\alpha}\exp\left(-\mu f_{\alpha}\right) \:,
\end{eqnarray}
with $\alpha$ a sum over pure states. By pure states we are referring to independent ergodic components in our system.  If we call
$P_{\alpha} = \exp\left(-\mu f_{\alpha}\right)/\left(\sum_{\alpha}\exp\left(-\mu
f_{\alpha}\right)\right)$ with $f_{\alpha}$ the free energy of state $\alpha$
we see that
\begin{eqnarray}
 \Sigma &\equiv& -\sum_{\alpha}P_{\alpha}\log\left(P_{\alpha}\right) = \mu(\Phi-f) \:,
\end{eqnarray}
with
\begin{eqnarray}
 -\beta f &=&  \sum_{\alpha}P_{\alpha}f_{\alpha}\:.
\end{eqnarray}
These quantities can be calculated through the following ansatz (see \cite{Mon1998}),
\begin{equation}
 \fl P(\bnu|\bsigma,a) = \int d\pi \, \mathcal{P}_{\rm 1RSB}\left(\pi|\bsigma,a\right) \prod^{\frac{n}{m}}_{\alpha=1} \left(\int dx \pi(x)\frac{\exp\left(\beta x \sum^m_{\gamma=1}\nu_{\alpha,\gamma}\right)}{\left(2\cosh\left(\beta x \right)\right)^m} \right)\:, \label{eq:1RSBNuA}
\end{equation}
for some functional $\mathcal{P}_{\rm 1RSB}[\pi|\bsigma,a]$
with $\int\mathcal{D}\pi\mathcal{P}_{\rm
1RSB}[\pi|\bsigma,a]=1$. Replicas here are only interchangeable
within the group $\alpha=1,\ldots,\frac{n}{m}$ to which they belong.
Spin variables carry two indices denoting the group $\alpha$ and
replica $\gamma$ within the group. This one-step replica symmetry
breaking has been considered for the binary symmetric channel in
\cite{Migl2005}.

Substituting this ansatz into the self-consistent equations
(\ref{eq:selfc}) results in, using $\beta m =\mu$,
\begin{eqnarray}
 \fl \mathcal{P}_{\rm 1RSB}\left(\pi|\bsigma,a\right) =  \prod^{C-1}_{r=1} \left(\frac{\int \prod_{l}\mathcal{D}_b\,a_l \sum_{\bsigma_1, \cdots, \bsigma_{K-1}}\delta\left(\bsigma\prod_{l}\bsigma_l;1\right)\prod_{l}P(\bsigma_l|a_l) }{\int \prod_{l}\mathcal{D}_b\,a'_l\left(\sum_{\bsigma'_1, \cdots, \bsigma'_{K-1}}\delta\left(\bsigma'\prod_{l}\bsigma'_l;1\right)\prod_{l}P(\bsigma'_l|a'_l)\right)} \right.
\nonumber \\  \times \left.\int\prod_l d\pi^r_l\mathcal{P}_{\rm
1RSB}\left(\pi^r_l|\bsigma_l, a_l\right)\right)\int dh
p(h|\sigma^1,a) \delta_F\Big[\pi(x) - \mathcal{U}(x;
\left\{\mathcal{\pi}^r_l\right\}, h)\Big]\:. \nonumber \\ 
\label{eq:1RSBselfcEq}
\end{eqnarray}
where $\delta_F[\xi(x)]$ denotes a functional delta distribution in
the sense that $\mathcal{Q}[f]=\int d\xi
\mathcal{Q}[\xi]\delta_F[\xi(x)-f(x)]$.  We also introduced the
distribution $\mathcal{U}(x; \left\{\mathcal{\pi}^r_l\right\}, h)$, equal to:
\begin{eqnarray}
 \fl \mathcal{U}(x; \left\{\mathcal{\pi}^r_l\right\}, h) = \int \prod^{C-1}_{r=1}\prod^{K-1}_{l=1} dx^r_l\pi^r_l(x^r_l)\exp\left(-\mu \Delta F\right)
 \delta\Big(x-u_{\beta}\left(\left\{x^r_l\right\},h\right)\Big)\:. \label{eq:U}
\end{eqnarray}
In (\ref{eq:U})
$\Delta F$ is given by
\begin{eqnarray}
 \Delta F &=& -\frac{1}{\beta}\log\left(\sum_{\tau}\exp\left(\beta h\tau\right)\prod^{C-1}_{r=1}\frac{1}{2}\left(1+\tau\prod^{K-1}_{l=1}\tanh(\beta x^r_l)\right)\right)\:.
\end{eqnarray}
which in terms of the cavity terminology equals the free energy
shift due to iteration. When we focus from now on on the unbiased case
we get a somewhat simpler expression:
\begin{eqnarray}
  \fl \mathcal{P}_{\rm 1RSB}\left(\pi|\sigma\right) = \prod^{C-1}_{r=1}\left( \sum_{\sigma_1, \sigma_2, \cdots, \sigma_{K-1}}\frac{\delta\left(\sigma\prod_{l}\sigma_l;1\right)}{2^{K-2}} \int\prod^{K-1}_{l=1}d\pi^r_l\mathcal{P}_{\rm 1RSB}\left(\pi^r_l|\sigma_l\right)\right)
\nonumber \\
\times \int dh \: p(h|\sigma)\: \delta_F\left[\pi(x) - \mathcal{U}(x; \left\{\mathcal{\pi}^r_l\right\}, h)\right] \:. \label{eq:1RSBun}
\end{eqnarray}
In principle this equation can be solved
with the iterative scheme of population dynamics \cite{Mez2001}.
Substitution of (\ref{eq:1RSBNuA}) and (\ref{eq:1RSBselfcEq}) in (\ref{eq:FreeE})
produces an expression for the generalized free energy $\Phi_{\rm
1RSB}(\mu)$
\begin{eqnarray}
 \fl-\Phi_{1RSB} =\left(\frac{C}{K}\left(K-1\right)\right) \mathbb{E}^{(K)}_{1RSB}\left[\Delta \Phi^{(K)}_{1RSB}\left(\left\{\pi_l\right\}\right)\right] - \mathbb{E}^{(1)}_{1RSB}\left[\Delta \Phi^{(1)}_{1RSB}\left(\left\{\pi^r_l\right\}; h\right)\right]\:,\label{eq:1rsbfree}
\end{eqnarray}
with the averages
\begin{eqnarray}
 \fl \mathbb{E}^{(K)}_{1RSB} \left[g\left(\left\{\pi_l\right\}\right)\right] =   \left(\sum_{\sigma_1,\cdots, \sigma_{K}}\frac{\delta\left(\prod_{l}\sigma_l\right)}{2^{K-1}} \int \prod^{K}_{l=1}\mathcal{D}\pi_l\: \mathcal{P}(\pi_l|\sigma_l)\right) g\left(\left\{\pi_l\right\}\right)\:,\\
\fl \mathbb{E}^{(1)}_{1RSB}\left[g(\left\{\pi^r_l\right\}; h)\right] =   \frac{1}{2}\sum_{\sigma} \left(\prod^{C}_{r=1}\sum_{\sigma^r_1\cdots\sigma^r_{K-1}}\frac{\delta\left(\sigma\prod_{l}\sigma^r_l;1\right)}{2^{K-2}}\right)
\nonumber \\
\times \int  \prod_{r,l} d\pi^r_l\mathcal{P}(\pi^r_l|\sigma^r_l)\int dh \: p(h|\sigma,\frac{1}{2}) g(\left\{\pi^r_l\right\}; h) \:.
\end{eqnarray}
The generalized free energy shifts  $\Delta \Phi_{\rm 1RSB}^{(K)}$ and $\Delta \Phi_{\rm 1RSB}^{(1)}$ are given by:
\begin{eqnarray}
 \Delta \Phi_{\rm 1RSB}^{(K)} &=& -\frac{1}{\mu}\log\left(\int \left(\prod^{K}_{l=1}dx_l \: \pi_l(x_l|\sigma_l)\right)\exp\left[-\mu \Delta F_{\rm RS}^{(K)}\right]\right) \:,\\
\Delta \Phi_{\rm 1RSB}^{(1)}&=& -\frac{1}{\mu}\log\left(\int
\left(\prod^{K-1}_{l=1}\prod^{C}_{r=1}dx^r_l \:
\pi^r_l\left(x^r_l|\sigma^r_l\right) \right) \exp\left[-\mu\Delta
F_{\rm RS}^{(1)}\right]\right)\:.
\end{eqnarray}
The free energy follows from $f_{\rm 1RSB}(\mu) =
\partial (\mu \Phi)/\partial \mu$:
\begin{eqnarray}
 \fl -f_{1RSB} =\left(\frac{C}{K}\left(K-1\right)\right) \mathbb{E}^{(K)}_{1RSB}\left[\Delta f^{(K)}_{1RSB}\left(\left\{\pi_l\right\}\right)\right] - \mathbb{E}^{(1)}_{1RSB}\left[\Delta f^{(1)}_{1RSB}\left(\left\{\pi^r_l\right\}; h\right)\right]\:,\label{eq:f1RSB}
\end{eqnarray}
with
\begin{eqnarray}
 \Delta f^{(K)}_{1RSB} &=& \frac{\int \left(\prod^K_{l=1}dx_l\pi_l(x_l|\sigma_l)\right) \Delta F_{\rm RS}^{(K)} \exp\left[-\mu \Delta F_{\rm RS}^{(K)}\right]}{\int \left(\prod^K_{l=1}dx_l\pi_l(x_l|\sigma_l)\right) \exp\left[-\mu \Delta F_{\rm RS}^{(K)}\right] } \:,
\\
\Delta f^{(1)}_{1RSB} &=& \frac{\int \left(\prod^{K-1}_{l=1}\prod^{C}_{r=1}dx^r_l
\pi^r_l\left(x^r_l|\sigma^r_l\right) \right)\Delta
F_{\rm RS}^{(1)}\exp\left[-\mu\Delta
F_{\rm RS}^{(1)}\right]}{\int
\left(\prod^{K-1}_{l=1}\prod^{C}_{r=1}dx^r_l
\pi^r_l\left(x^r_l|\sigma^r_l\right) \right)\exp\left[-\mu\Delta
F_{\rm RS}^{(1)}\right]} \:.
\end{eqnarray}
Combining (\ref{eq:1rsbfree}) and (\ref{eq:f1RSB}) produces finally the
complexity: 
\begin{eqnarray}
 \Sigma(f_{\rm 1RSB}) = \mu f_{1RSB} - \mu \Phi(\mu) \:, \label{eq:f*}
\end{eqnarray}
as a function of the free energy. An alternative way to derive the
above is based on the cavity method (see \ref{sec:cav})
which shows that the complexity corresponds to the entropy of the
number of solutions to the cavity equations with free energy density
$f_{\rm 1RSB}$.

\begin{figure}
\begin{minipage}{.45\textwidth}
\begin{center}
\includegraphics[angle=-90, width = 1 \textwidth]{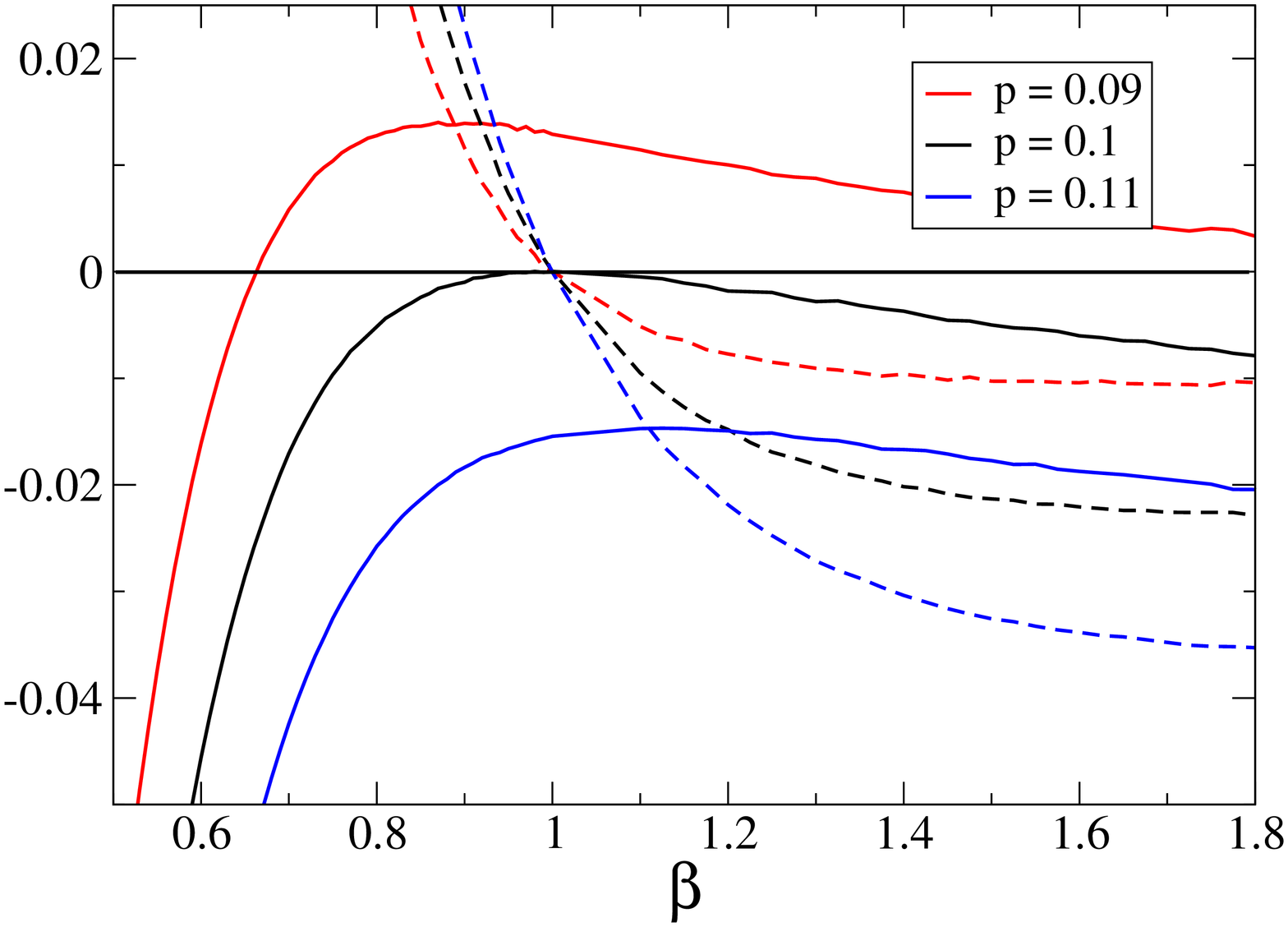}
\end{center}
\end{minipage}
\hfill
\begin{minipage}{.45\textwidth}
\begin{center}
\includegraphics[angle=-90,width= 1 \textwidth]{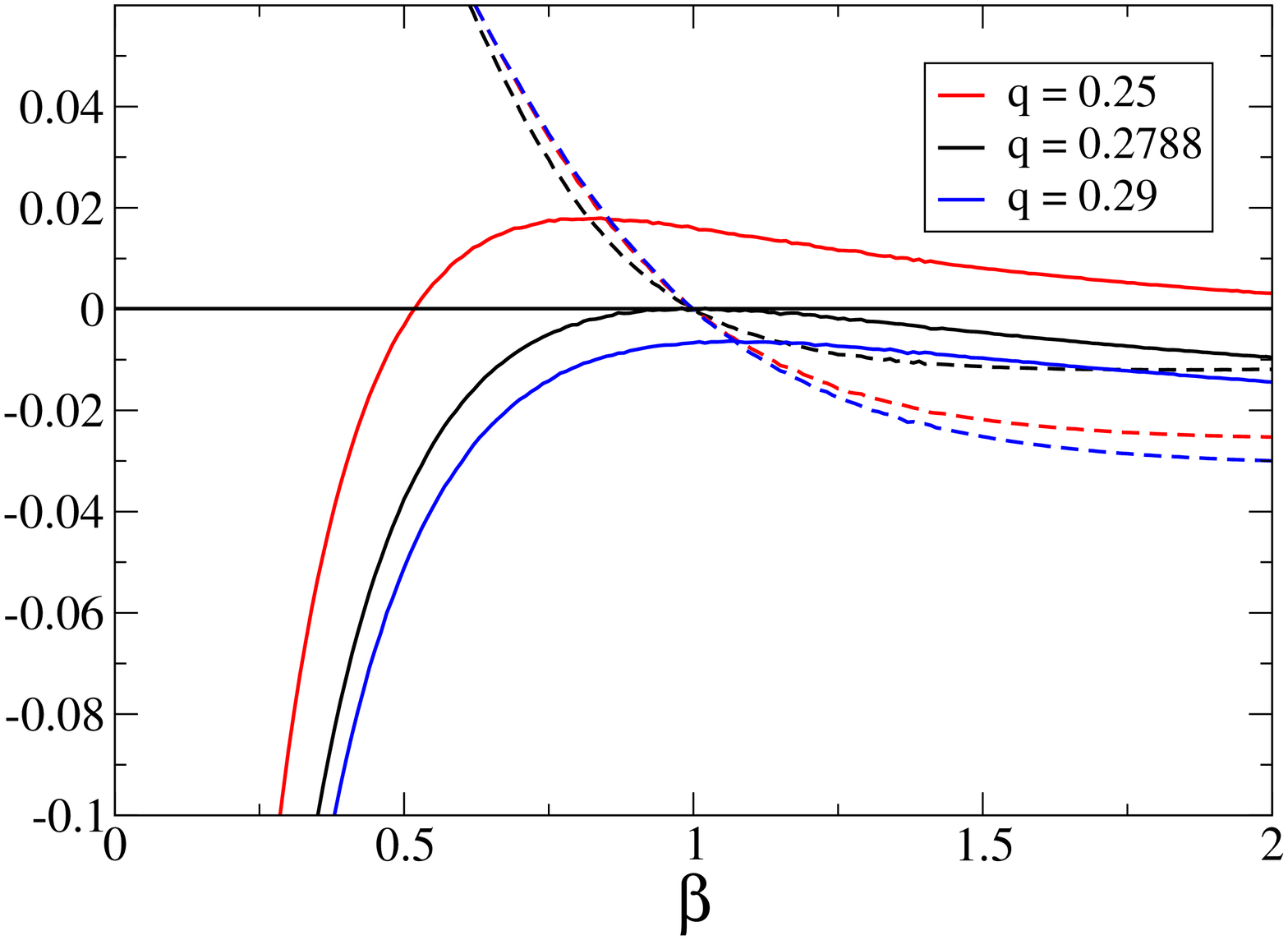}
\end{center}
\end{minipage}\caption{The free energy $f-e_{\rm ferro}$ (solid line)
and the energy difference $e-e_{\rm ferro}$
(dashed line) as a function of the inverse temperature $\beta$
for a regular $(C,K)=(3,6)$ Gallager code.   At the point $f=e$ the entropy is zero. This point
determines the thermodynamic value of $f$ at the frozen transition.
These graphs can also be interpreted in the frozen ansatz (\ref{eq:codeword}) in 1RSB, with
the identification (\ref{eq:equivalences}).
Left: $\kappa=1$ (binary symmetric channel) with
$p=\{0.09,0.1,0.11\}$ from top to bottom. Right: $\kappa=0$
(Z-channel) with $q=\{0.25,0.2788,0.29\}$ from top to bottom.}
\label{fig:AF1}
\end{figure}

\begin{figure}
\begin{minipage}{.45\textwidth}
\begin{center}
\includegraphics[angle=-90, width = 1 \textwidth]{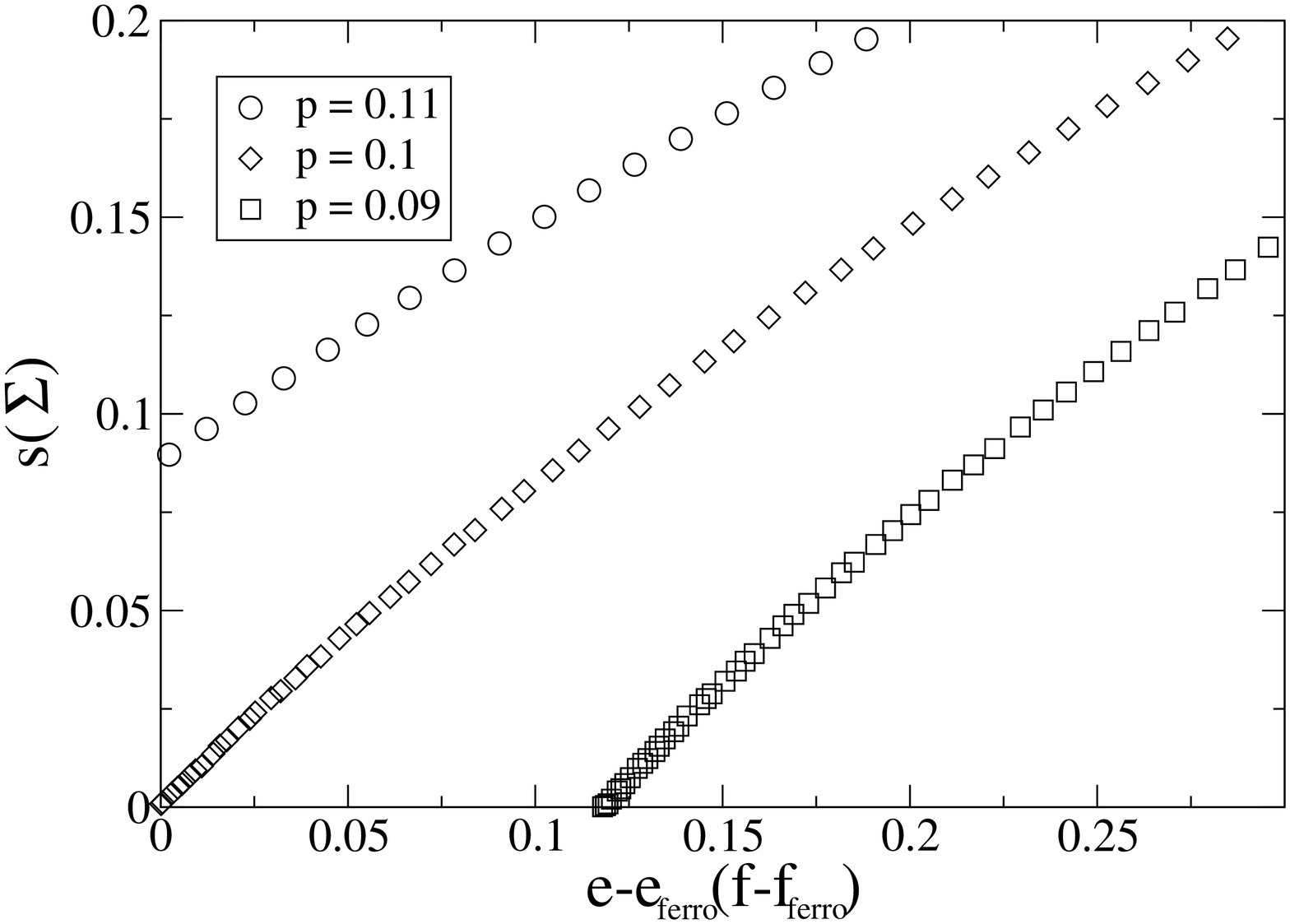}
\end{center}
\end{minipage}
\hfill
\begin{minipage}{.45\textwidth}
\begin{center}
\includegraphics[angle=-90,width= 1 \textwidth]{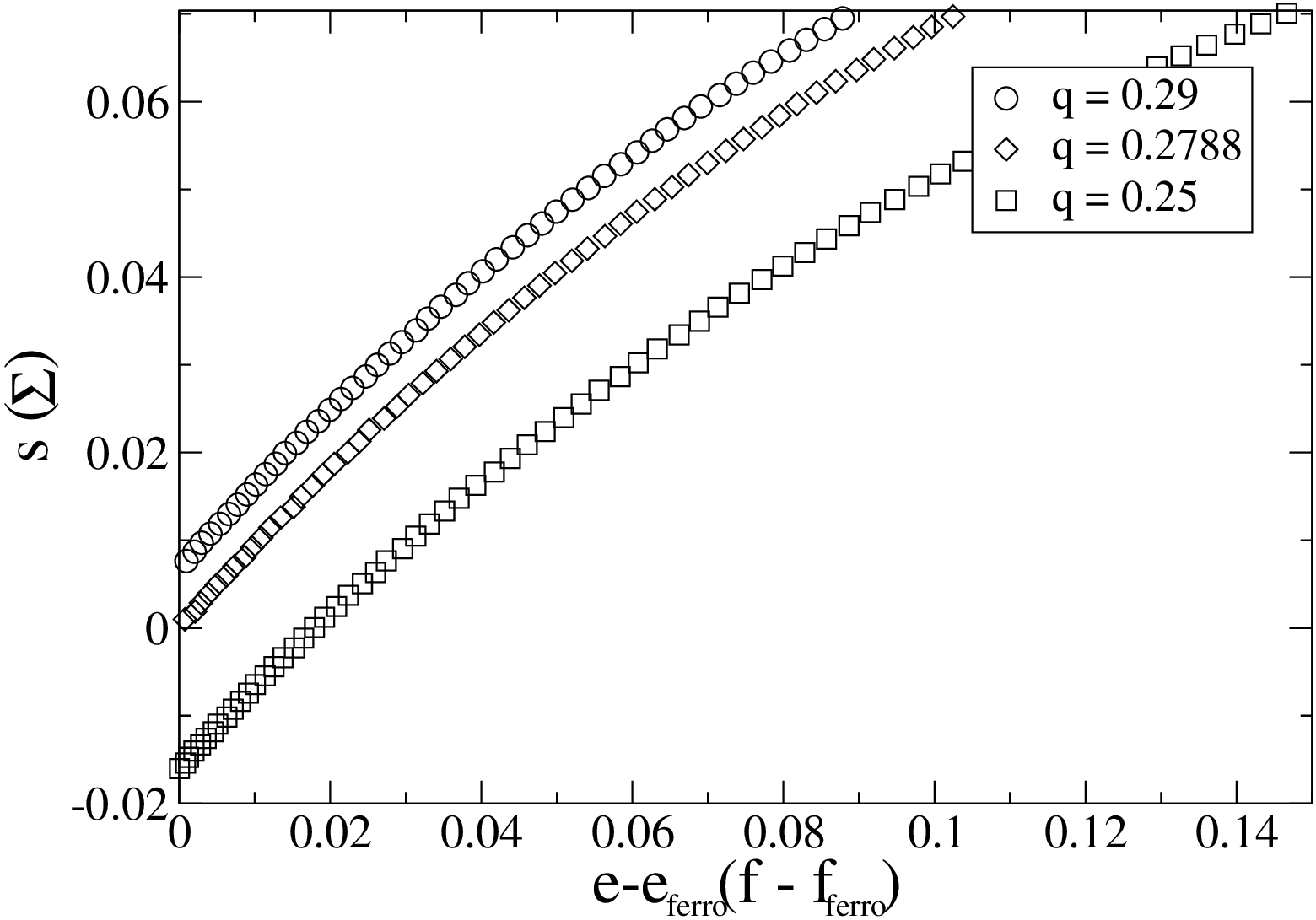}
\end{center}
\end{minipage}
\caption{The entropy $s$ (complexity $\Sigma$) of a regular $(C,K)=(3,6)$ Gallager code
as a function of the energy difference $e_{\rm RS}-e_{\rm ferro}$ (free energy difference $f_{\rm 1RSB}-e_{\rm ferro}$).
Left: binary symmetric channel, right: Z-channel.
Above the thermodynamic transition (upper line) the complexity
$\Sigma(f_{\rm ferro})$ is positive implying that there are
exponentially many codewords with the same free energy and as a
result decoding fails.} \label{fig:AF2}
\end{figure}

\subsection{A special case: the frozen-ansatz } \label{sec:frozen}
The general scheme described by (\ref{eq:1RSBselfcEq}) allows
one to retrieve the solutions of the frozen ansatz
(\ref{eq:frozenansatz}).
This can be done
by considering solutions of the form
\begin{eqnarray}
 \mathcal{P}[\pi|\sigma] &=& \int da\,Q(a|\sigma)\ \delta_F\Big(\pi(x) - a\delta\left(x-\infty\right) - (1-a)\delta\left(x+\infty\right)\Big) \:. \label{eq:codeword}
\end{eqnarray}
This solution can be interpreted as having on each site a probability $a$ to have a state with $\nu=1$.   The distribution $Q(a|\sigma)$ corresponds to site averages.
In the case where $C$ is even, it is clear that (\ref{eq:codeword}) is a solution of the 1RSB self-consistent equations (\ref{eq:1RSBselfcEq}).
For odd $C$ the reweighting factor $e^{-\mu \Delta F}$ in (\ref{eq:1RSBselfcEq}) makes sure that the zero fields do not appear.  Hence (\ref{eq:codeword})
is also a solution of (\ref{eq:1RSBselfcEq}) when $C$ is odd.
Substitution of (\ref{eq:codeword}) in  (\ref{eq:1RSBselfcEq}) gives the following self-consistent equation for the distribution $Q(a)$
\begin{eqnarray}
 \fl Q(a|\sigma) = \prod^{C-1}_{r=1}\left( \sum_{\sigma_1, \sigma_2, \cdots, \sigma_{K-1}}\frac{\delta\left(\sigma\prod_{l}\sigma_l;1\right)}{2^{K-2}} \int\prod^{K-1}_{l=1}da^r_lQ\left(a^r_l|\sigma_l\right)\right)
\nonumber \\
 \times
\int dh\, p(h|\sigma) \int d\tilde{a}_+d\tilde{a}_-\int d\mathcal{N} \delta\left(\mathcal{N}-\left(\tilde{a}_++\tilde{a}_-\right)\right)
\nonumber \\
\delta \left[a  - \frac{\exp\left(\mu h\right)}{\mathcal{N}}\prod^{C-1}_{r=1}\left[\sum_{\substack{n=0 \ n\, {\rm is\, even}}}^{K-1}  \sum_{(l_1,  \cdots, l_n)}\prod^n_{i=1}\left(1- a^{r}_{l_i}\right)\prod_{l\notin (l_1,  \cdots, l_n)}a^{r}_{l}\right]\right]
\nonumber \\
\delta \left[\tilde{a}_+  - \exp\left(\mu h\right)\prod^{C-1}_{r=1}\left[\sum_{\substack{n=0 \ n\, {\rm is\, even}}}^{K-1}\sum_{(l_1,  \cdots, l_n)}\prod^n_{i=1}\left(1- a^{r}_{l_i}\right)\prod_{l\notin (l_1, \cdots, l_n)}a^{r}_{l}\right]\right]
\nonumber \\
 \delta \left[\tilde{a}_-- \exp\left(-\mu h\right) \prod^{C-1}_{r=1}\left[\sum_{\substack{n=0 \ n\, {\rm is\, odd}}}^{K-1}\sum_{(l_1,  \cdots, l_n)}\prod^n_{i=1}\left(1- a^{r}_{l_i}\right)\prod_{l\notin (l_1, \cdots, l_n)}a^{r}_{l}\right]\right]  \:.
\end{eqnarray}
These equations turn out to be equivalent to the RS equations (\ref{eq:distriBACdistri}), we found before. This can be seen by substituting
$a =\exp\left(\mu x\right)/\left(2\cosh\left(\beta x\right)\right)$. Then setting $\mu = \beta$ we obtain the following identities between RS and the present ansatz (\ref{eq:codeword}):
\begin{equation}
\Sigma_{\rm 1RSB} = s_{\rm RS}\:,
\hspace{7mm}
\Phi_{\rm 1RSB} = f_{\rm RS}\:,
\hspace{7mm}
f_{\rm 1RSB} = \epsilon_{\rm RS}\:. \label{eq:equivalences}
\end{equation}
We have thus returned to replica symmetry with the 1RSB free energy
playing the role of the RS energy.  From figure \ref{fig:AF1} we
see that indeed $f_{RS}(\Phi_{1RSB})$ reaches its maximum at
$s_{RS} = 0 \:(\Sigma_{1RSB} = 0)$.  From this it follows that we can
maximize the free energy for $\beta <\beta_g$ with $m=1$, while for
$\beta>\beta_g$  we obtain $m=\beta_g/\beta$. The relationships
(\ref{eq:equivalences}) can easily be interpreted through:
\begin{eqnarray}
 \fl-\mu \Phi_{1RSB}(\mu) &=  \log\left(\sum_{{\rm states}\ \alpha}\exp\left(-\mu f_{\alpha}\right)\right)
\nonumber \\ 
 &= \log\left(\sum_{{\rm states}\ \alpha}\exp\left(- \mu \epsilon_{\alpha}\right)\right) = -\beta mf_{RS}(\beta m)\:,
\end{eqnarray}
where we used that $s_{\alpha} = 0$.  In figure \ref{fig:AF2} we see that above the thermodynamical noise levels $(p_c,q_c)$, the number of codewords with an energy equal to the ferromagnetic codeword scales exponentially with the system size.  These figures can be compared with the figures \ref{fig:shannon}.  Just like in the RCM we can interpret the thermodynamical transition $(p_c,q_c)$ at $T=1$ as the theoretical upper limit for succesful decoding with (3,6)-Gallager codes.

\subsection{The more general case: the complete 1RSB}

\begin{figure}[h!dt]
\begin{center}
\includegraphics[angle=-90, width=.6 \textwidth]{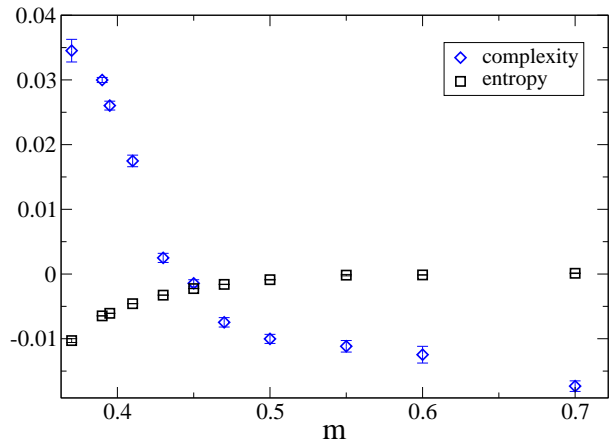}
\caption{The complexity $\Sigma$ and the entropy $s$ as a
function of the replica symmetry breaking parameter $m$ for a BSC
and a $(3,6)$-code at $p=0.09$ and $T=\frac{1}{2}$, which
corresponds to the marked point ($\bullet$) in the phase diagram of
figure \ref{fig:kappaFull}.}\label{fig:comple1RSB}
\end{center} 
\end{figure}

Finally we consider the  solution of the 1RSB equations (\ref{eq:1RSBun}) and (\ref{eq:1rsbfree}). Numerically
the 1RSB approach is prone to many errors coming from the small
sizes of the distributions (we used 1000 distributions of each 1000
fields).  We also emphasize that the 1RSB replica and cavity method
may contain many non controllable approximations.  This is especially true for the complexity, see
\cite{Mont2003} and \cite{Cav2005}.  With this in mind we try to
interpret the result presented in figure
\ref{fig:comple1RSB}, which has been calculated for parameter values
corresponding to the marked points of the phase diagram in
figure \ref{fig:kappaFull}. At the point marked with a cross we find a zero complexity for all values of $m$.  At the dotted marker we find a positive complexity for some values of $m$. From the thermodynamical relation between $m$ and $\Sigma$, we know that $\Sigma$ must decrease as a function of $m$.  Eliminating the branches where the complexity increases as a function of $m$ we find the results in figure \ref{fig:comple1RSB}.  We find a regime with a positive complexity and a negative entropy.  This means that there is an exponential number of solutions to the belief propagation equations and thus the belief propagation algorithm does no longer converge.   We also remark that the fact that these solutions have a negative entropy is consistent with the freezing picture we found in section \ref{sec:frozen}.  It would be
interesting to look for the change of the dynamic thresholds between
the replica symmetric and 1RSB algorithms.   In order to exclude
finite size effects, we would need larger system sizes to determine
accurately these thresholds.

\section{Discussion}
\label{sec:disc}

In this paper we study the decoding properties of LDPC-codes on a binary asymmetric channel, using
tools from statistical mechanics on finitely connected systems. As a result of the channel asymmetry the microscopic
Boltzmann distribution for the channel noise
inherits an explicit dependence on the received message. This results in a set of recursive equations
for two types of cavity fields.  We determine the decoding thresholds for message passing algorithms as a function of the important parameters, e.g. the asymmetry, the bias and the temperature.  Calculating the entropy we find the upper bound to any decoding scheme.

For dense codes we retrieve the random codeword model.  The thermodynamic averages are characterized by the existence of a ferromagnetic, spin glass and paramagnetic phase. The ferromagnetic region increases with increasing asymmetry in the channel noise.  Because the paramagnetic solution is always stable, the message passing algorithms fail to decode  correctly the received message at all noise levels. 
 
For low-density codes the emerging picture in the temperature-noise phase diagram is that for high temperatures we find two solutions for the cavity distributions.   For these temperatures and low noise levels there is a ferromagnetic phase indicating succesful decoding.  Increasing the noise level to a certain threshold the appearance of a paramagnetic solution distorts the decoding process.  As the temperature is lowered this paramagnetic solution freezes into a zero-entropy solution, representing a subexponential number of codewords.  Lowering the temperature even further the decoding dynamics of the system is distorted by an exponential number of metastable states.  We discuss this failure in terms of the endogeny property of the recursive equations for the cavity fields.

\newpage

\appendix

\section{The capacity of the binary asymmetric channel}\label{app:capBAC}

\begin{figure}
\begin{center}
\includegraphics[angle=-90, width=.5 \textwidth]{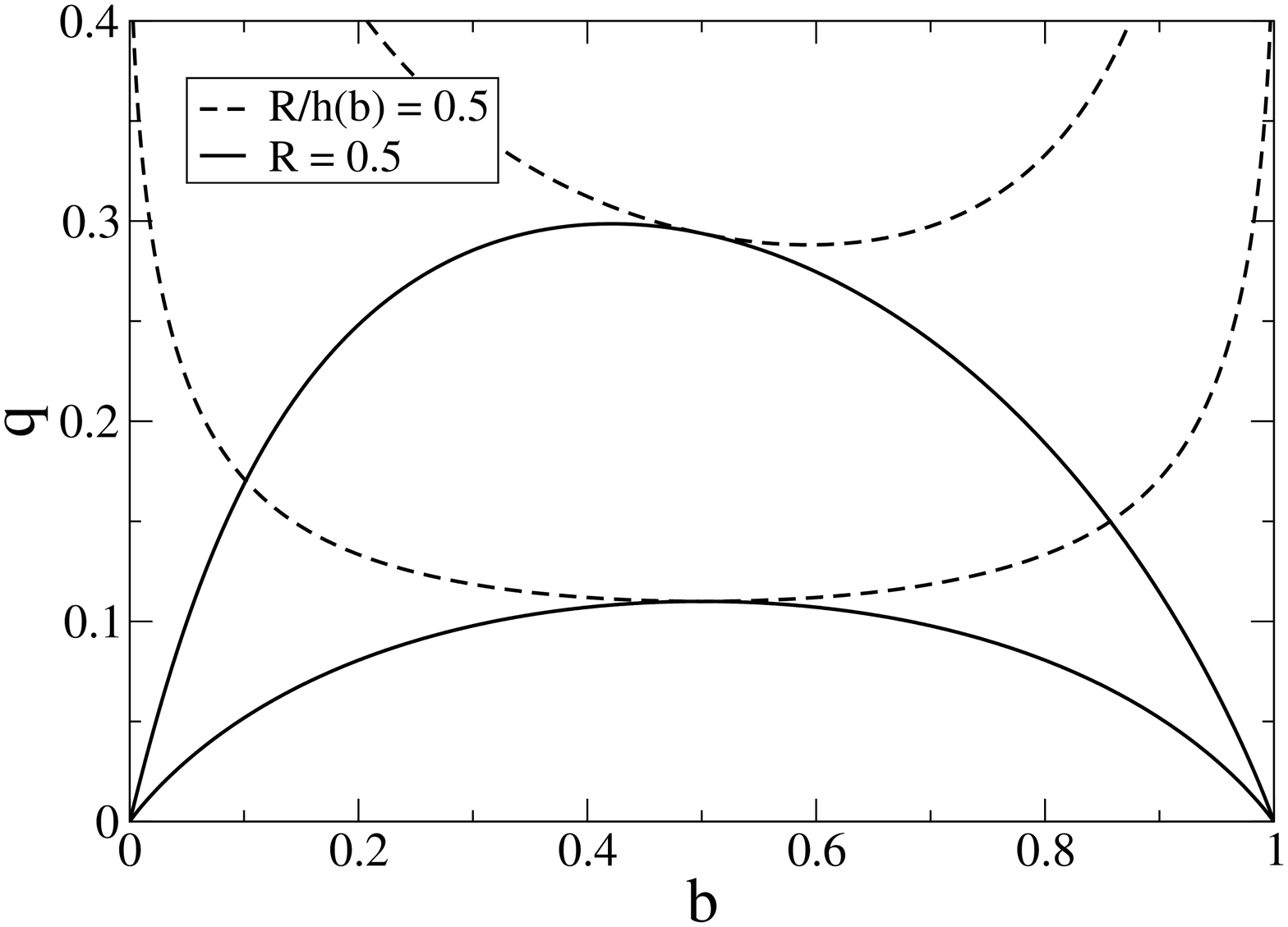}
\caption{The Shannon limit of the noise $q$ as a function of the bias $b$, given the rate R or the fraction $R/h(b)$.  
The curves symmetric with respect to $b=\frac{1}{2}$ correspond to a BSC channel.  The other curves represent a Z-channel.  
The solid lines are calculated for a biased source and the dashed lines for a biased source with unbiased redundant bits.  \label{fig:shannonLimB}}
\end{center}
\end{figure}

Shannon's famous channel coding theorem states that error free
communication can be possible as long as the rate $R = \frac{N}{M}h(b)$
is kept below a certain critical value $\mathtt{C}$, the
channel capacity. We use the abbreviation $h(b)$ for the binary entropy,
$h(t)\equiv -t\log_2 t-(1-t)\log_2(1-t)$.
We here calculate the channel capacity for
the binary asymmetric channel. It is defined as
\begin{equation}
\mathtt{C}=\underset{p(X)}{\rm max}\ \mathcal{I}(X,Y) \:,
\label{eq:capacity}
\end{equation}
where $X=\left\{x_1, x_2, \cdots, x_L\right\}$ is the set of
possible inputs to the channel and $Y = \left\{y_1,  y_2,\cdots,
y_{\tilde{L}}\right\}$ the set of possible outputs. The average
mutual information
\begin{equation}
\mathcal{I}(X,Y)=H(Y)-H(Y|X) \label{eq:mutualinfo}\:,
\end{equation}
written in terms of the marginal- and conditional entropy
represents the amount of information carried by the channel for a
given noise probability. Thus $\mathtt{C}$ provides the maximum
admissible amount of information carried by the channel.
In
the case of a binary alphabet with $L=\tilde{L}=2$ and with $
p(X)=b\delta_{X,x_1}+(1-b)\delta_{X,x_2}$ we find that for the
binary asymmetric channel, see figure \ref{fig:channel}, the mutual information is given by
\begin{equation}
\mathcal{I}(p,q) = h\left(\frac{b(1-q)+(1-b)p}{2}\right) - \frac{b h(q) + (1-b) h(p)}{2} \:. \label{eq:mutualIas}
\end{equation}
The maximum of
$\mathcal{I}(p,q)$ with respect to $b$ is attained at
\begin{equation}
b_{\star}=\frac{p\left(\exp\left[F(p,q)\right]+1\right)-1}
{(q+p-1)\left(1+\exp\left[F(p,q)\right]\right)} \:,
\end{equation}
with $F(p,q)=[\frac{h(p)-h(q)}{q+p-1}]\log 2$, provided that
$b_{\star}\neq p/(p+q-1)$.  Indeed, we see in figure (\ref{fig:shannonLimB}) that at constant rate the channel noise gets a maximum at some $b\neq \frac{1}{2}$.  If on the other hand we keep the code $(C,K)$ fixed, and we take into consideration that the parity check bits are unbiased, we get a minimum value at $b\neq \frac{1}{2}$.

\section{Cavity method} \label{sec:cav}
We can derive mean field equations for a specific graph
instance using the cavity method \cite{meZ2003}.  The cavity method
gives us a link between the different mean field solutions we find
using the replica method and different decoding algorithms.
First, we derive the cavity equations for a typical solution
$\bsigma$ with a weight given by (\ref{eq:PriorSigma}). We define
the cavity graph $\mathcal{G}_{M, q}$ as a graph having $M$
spins connected to $C$ hyperedges and $q$ cavity spins
connected to $C-1$ hyperedges.  On this graph we consider the
graph operations defined in \cite{meZ2003}: site addition, link
addition and site iteration.
 We will try to count how the number of solutions $\mathcal{N}(e)$  to the equations,
\begin{eqnarray}
 \delta\left(\prod_{i\in\omega_j}\nu_i; 1\right) = 1\:,\ j=1,2,\cdots,M-N \:,\label{eq: phaseSpa}
\end{eqnarray}
corresponding with an energy density $e = \frac{E}{M}$, change when
performing the aforementioned graph operations. If
$\mathcal{C}$ is the set of solutions of (\ref{eq: phaseSpa}), we
define
\begin{eqnarray}
 \mathcal{N}(e) &=& \#\left\{\bnu \in \mathcal{C}: \frac{E(\bnu)}{M} = -\frac{\sum_{i}\nu_ih_i}{M} = e\right\} \sim \exp\left(M s(e)\right) \:,
\end{eqnarray}
$s(e)$ is the entropy and $h_i$ are the quenched fields.  Suppose we are only interested in small fluctuations around some reference energy $E_{\rm ref}$.  The probability $P(E)$ that a configuration has an energy $E$ is then given by
\begin{eqnarray}
 P(E) \sim \exp\left(M s\left(\frac{E_{\rm ref}+\Delta E}{M}\right)\right) \sim \exp\left(\beta \Delta E\right) \:,
\end{eqnarray}
with $\beta = \frac{\partial s}{\partial e}$.   We define through a Legendre transform the free energy F
\begin{eqnarray}
 \beta F(\beta) &=& M\left(\beta e - s(e)\right)\:.
\end{eqnarray}
We use the notation $s_{M, q}$ for the entropy density on the graph $\mathcal{G}_{M, q}$. Site addition is the graph operation which adds a site to $\mathcal{G}_{M, q}$ connecting it with $C$ hyperedges to $C(K-1)$ cavity spins.  Under site addition $\mathcal{N}\left(e\right)$, we assume that $\mathcal{N}\left(e\right)$ fullfills
\begin{eqnarray}
\fl \exp\left(\left(M+1\right)s_{M+1, 0}\left(\frac{E}{M+1}\right)\right)
\nonumber \\
= \int P^{(E)}_{site}(\Delta E)\exp\left(M s_{M, C(K-1)}\left(\frac{E-\Delta E}{M}\right)\right)d\Delta E
\nonumber \\
= \exp\left(M s_{M, C(K-1)}\left(\frac{E}{M}\right)\right) \int d\Delta E P^{(E)}_{site}(\Delta E)\exp\left(-\beta \Delta E\right)\:, \label{eq:siteAde}
\end{eqnarray}
 where $P^{(E)}_{site}(\Delta E)$ is the distribution of energy changes under site addition.
From (\ref{eq:siteAde}) we have
\begin{eqnarray}
\fl \exp\left[(M+1)s_{M+1, 0}(e)-Ms_{M, C(K-1)}(e) - \beta e\right]
\nonumber \\
=\exp\left[\log\left(\int d\Delta P^{(E)}_{site}\left(\Delta E\right)\exp\left(-\beta \Delta E\right)\right)\right]\:.
\end{eqnarray}
The free energy change under site addition, $\Delta F^{(1)}$, is thus equal to
\begin{eqnarray}
 \Delta F^{(1)} &=& -\frac{1}{\beta}\log\left(\int d\Delta E P^{(E)}_{site}\left(\Delta E\right)\exp\left(-\beta \Delta E\right)\right)\:. \label{eq:DF1}
\end{eqnarray}
Link addition is the graph operation which adds a hyperedge between K cavity spins. Under this operation we find for $\mathcal{N}\left(e\right)$
\begin{eqnarray}
 \exp\left(Ms_{M, 0}(e)\right) &=& \exp\left(Ms_{M, K}(e)\right) \left(\int d\Delta E P^{(E)}_{link}\left(\Delta E\right) \exp\left(-\beta \Delta E\right) \right)  \:.\nonumber \\
\end{eqnarray}
The free energy change under link addition $\Delta F^{(K)}$ becomes
\begin{eqnarray}
 \Delta F^{(K)} &=&  -\frac{1}{\beta}\log\left(\int d\Delta E P^{(E)}_{link}\left(\Delta E\right)\exp\left(-\beta \Delta E\right)\right)\label{eq:DFK} \:.
\end{eqnarray}
When we start from a graph  $\mathcal{G}_{M, CK(K-1))}$, we can perform K site additions or $C(K-1)$ link additions to get a graph without cavity spins.  In the limit $M\rightarrow \infty$ we get
\begin{eqnarray}
 F &=&  \Delta F^{(1)} - \frac{C(K-1)}{K}\Delta F^{(K)}\:. \label{eq:Frec}
\end{eqnarray}
The distributions  $P_{link}\left(\Delta E\right)$ and $P_{site}\left(\Delta E\right)$ are given by
\begin{eqnarray}
 \fl P^{(E)}_{site}(\Delta E) = \sum_{\sigma_0}\sum_{\sigma_{1,1}, \sigma_{1,2},\cdots, \sigma_{K-1, C}} \prod^{C}_{r=1}\left[\prod^{K-1}_{l=1}P^{(E)}_{r,l}(\sigma_{r,l}) \delta\left(\sigma_0\prod^{K-1}_{l=1}\sigma_{r,l}; 1\right)\right]  \delta\left(\Delta E + h_0\nu_0  \right)\:, \nonumber \\
\fl P^{(E)}_{link}\left(\Delta E\right) =\sum_{\nu_1,\nu_2, \cdots, \nu_K} \left(\prod^{K}_{l=1}P^{(E)}_{l}(\nu_{l})\right) \delta\left(\prod^K_{l=1}\nu_l; 1\right) \delta\left(\Delta E \right) \:,
\end{eqnarray}
with $h_0$ the external field at the new site. $P^{(E)}_r(\nu_r)$ is
the distribution of the spins on site $r$ when we go to a state with
energy E.  We assumed that the probabilities of the cavity spins are
uncorrelated.  It is possible to find a recursion relation for
$P^{(E)}_r(\nu_r)$  through
\begin{eqnarray}
 \fl P_0(\nu_0, \Delta E) = \sum_{\nu_1, \cdots, \nu_K}\left(\prod^{C-1}_{r=1}\prod^{K-1}_{l=1}P_{r,l}(\nu_{r,l})\right) \prod^{C-1}_{r=1}
\delta\left(\nu_0;\prod^{K-1}_{l=1}\nu_{r,l}\right)\delta \left(\Delta E+h_0\nu_0\right)\:.
\end{eqnarray}
The joint probability of the spin $\nu_0$ at the new site, and the energy $E'$ after iteration $R_0\left(\nu_0, E'_0\right)$  is
\begin{eqnarray}
 \fl R_0\left(\nu_0, E'_0\right) = \int dE_0d\Delta E_0 \exp\left(\beta(E_0-E_{ref})\right) P_0(\nu_0, \Delta E_0)\delta \left(E'_0-E_0-\Delta E_0\right) \nonumber \\
\sim\exp\left(\beta(E'_0-E'_{ref})\right) P^{(E_0)}_0(\nu_0)\:,
\end{eqnarray}
with
\begin{eqnarray}
\fl P^{(E_0)}_0(\nu_0) =  \sum_{\nu_{1,1}, \cdots, \nu_{C-1,K-1}}\left(\prod^{C-1}_{r=1}\prod^{K-1}_{l=1}P^{(E_0)}_{r,l}(\nu_{r,l})\right) \prod^{C-1}_{r=1}\delta\left(\nu_0;\prod^{K-1}_{l=1}\nu_{r,l}\right)\exp\left(\beta h_0\nu_0\right)\:.
\end{eqnarray}
We see that the $E_0$ dependency disappears.
We can parametrize the spin distributions $P_{r,l}(\nu_{r,l})$ as
\begin{eqnarray}
 P_{r,l}(\nu_{r,l}) &=& \frac{\exp\left(\beta x^r_l\nu_{r,l}\right)}{2\cosh\left(\beta h^r_l\right)}\:,
\end{eqnarray}
to get the cavity or belief propagation equations
\begin{eqnarray}
 x_0 &=& h_0 + \frac{1}{\beta}\sum^{C-1}_{r=1}\atanh\left(\prod^{K-1}_{l=1}\tanh\left(\beta x^r_l\right)\right)\:. \label{eq:cav2}
\end{eqnarray}
From (\ref{eq:cav2}) we can retrieve the equations (\ref{eq:selfcDistriFinal}), using
\begin{eqnarray}
 \fl \pi(x|\sigma,z,a) &=& \frac{1}{M}\sum^{M}_{i=1}\sum_{a\in \partial_i} \delta\left(x-x_{a\rightarrow i}\right) \delta\left(z-z_{a\rightarrow i}\right)\delta\left(\sigma-\sigma_i\right) \delta\left(a-a_i\right) \:,
\end{eqnarray}
and the assumption that we have large loops in the graph.  
We find for the free energy changes $\Delta F^{(1)}$ and $\Delta F^{(K)}$
\begin{eqnarray}
\fl \Delta F^{(1)} = -\frac{1}{\beta} \log\left(\sum_{\nu_0}\exp\left(\beta h_0 \nu_0\right)\prod^C_{r=1}\sum_{\nu_1, \nu_2, \cdots, \nu_{K-1} }\delta\left(\nu_0\prod^{K-1}_{l=1}\nu_{l}; 1\right)\exp\left(\beta \sum^{K-1}_{l=1}x_{r,l}\nu_l\right)\right)\:, \nonumber \\
\\
\fl\Delta F^{(K)} =-\frac{1}{\beta} \log\left( \sum_{\nu_1, \nu_2, \cdots, \nu_{K}}\delta\left(\prod^K_{l=1}\nu_l;1\right)\exp\left(\beta\sum^{K}_{l=1} x_l\nu_l\right)\right)\:.
\end{eqnarray}
Sometimes the cavity equations (\ref{eq:cav2}) do not converge
because there are many solutions to these equations and each part of
the graph converges to different kind of solutions.  The reason is 
that the cavity spins are not uncorrelated.  In that case we assume that
there are $\mathcal{M}(f) $ solutions to the cavity equations with
free energy f, i.e.\@
\begin{eqnarray}
  \mathcal{M}(f) &=& \#\left\{\bh \in \mathcal{S}_{\beta}: \frac{F(\bx)}{M} = f\right\} \sim \exp\left(M \Sigma(f)\right) \:,
\end{eqnarray}
We then find through a complete analogue calculation as above the 1RSB equations on a specific graph instance.   In that case $\mu = \frac{\partial \Sigma}{\partial f}$.

\section*{References}
\bibliographystyle{ieeetr} 
\bibliography{bibliography}

\end{document}